\documentclass[12pt]{article}

\usepackage{amssymb,amsmath,amsfonts,eurosym,geometry,ulem, color,setspace,sectsty,comment,footmisc,caption,pdflscape,array,hyperref, graphicx}
\usepackage{booktabs,siunitx, makecell}
\usepackage{natbib}
\usepackage{mathtools}
\usepackage{breqn}
\usepackage{cellspace}
\usepackage{makecell}
\usepackage{chngcntr}
\usepackage{adjustbox}
\usepackage{rotating}

\DeclareMathOperator*{\argmin}{arg\,min}

\usepackage[flushleft]{threeparttable}

\usepackage{dcolumn}
\newcolumntype{d}[1]{D..{#1}}

\usepackage{subcaption}

\usepackage{arydshln} 
\usepackage[T1]{fontenc}
\usepackage{tgpagella}

\hypersetup{
    colorlinks=true,
    linkcolor=blue,
	citecolor=red,
	urlcolor=red
}

\usepackage{booktabs}
\usepackage{siunitx}
\newcolumntype{d}{S[
    input-open-uncertainty=,
    input-close-uncertainty=,
    parse-numbers = false,
    table-align-text-pre=false,
    table-align-text-post=false
 ]}

\newcolumntype{L}[1]{>{\raggedright\let\newline\\arraybackslash\hspace{0pt}}m{#1}}
\newcolumntype{C}[1]{>{\centering\let\newline\\arraybackslash\hspace{0pt}}m{#1}}
\newcolumntype{R}[1]{>{\raggedleft\let\newline\\arraybackslash\hspace{0pt}}m{#1}}

\begin{document}

\begin{titlepage}
\title{The Mariana Environmental Disaster and its Labor Market Effects}
\author{Hugo Sant'Anna\thanks{University of Georgia - hsantanna@uga.edu}}
\date{\today}
\maketitle
\begin{abstract}
\noindent 
This paper examines the labor market impacts of the 2015 Mariana Dam disaster in Brazil. It contrasts two theoretical models: an urban spatial equilibrium model and a factor of production model, with diverging perspectives on environmental influences on labor outcomes. Utilizing rich national administrative and spatial data, the study reveals that the unusual environmental alteration, with minimal human capital loss, primarily affected outcomes via the factor of production channel. Nevertheless, spatial equilibrium dynamics are discernible within certain market segments. This research contributes to the growing literature on environmental changes and its economic consequences.\\
\vspace{0in}\\
\noindent\textbf{Keywords:} Enrivonmental Disasters, Spatial Equilibrium, Factors of Production\\
\vspace{0in}\\
\noindent\textbf{JEL Codes:} Q54, E24, O13\\

\bigskip
\end{abstract}
\setcounter{page}{0}
\thispagestyle{empty}
\end{titlepage}
\pagebreak \newpage

\doublespacing

\section{Introduction} \label{sec:introduction}

In the recent past, economies across the globe have witnessed severe disruption due to environmental disasters of both natural and anthropogenic origin. These calamities often cause widespread, long-term changes in the affected regions, leaving indelible marks on their landscapes, altering spatial equilibrium, and impacting various economic dimensions, from local labor markets to international trade. Disentangling the multifaceted effects of such incidents is challenging, especially given the vast heterogeneity across such disasters and their consequences. 

This study exploits the unique circumstances surrounding the Mariana Dam disaster in Brazil to understand these interconnected effects. A key aspect setting apart the Mariana disaster from other similar incidents is its geographically concentrated impact but extensive spillover. The disaster's epicenter was Bento Rodrigues, a small district in the municipality of Mariana, Minas Gerais, Brazil, with a population of 600 inhabitants. The impending dam rupture prompted a swift relocation of these residents, averting a potential human catastrophe on a larger scale. Therefore, only 19 fatalities occurred despite the total submersion of Bento Rodrigues. Nevertheless, the aftermath of the Mariana Dam rupture saw the severe contamination of the Doce River, a vital water source and livelihood provider for approximately two million individuals in Minas Gerais and Espírito Santo states. By December 2015, around 55 million cubic meters of toxic iron tailing waste had flowed downstream in a 663.2 km course, contaminating the river ecosystem and its 80 km radius surroundings before reaching the Atlantic Ocean in Espírito Santo. This ecological disaster has been tagged the worst in Brazilian history and the most severe globally involving mining operations. The peculiar nature of the Mariana disaster - immense environmental devastation alongside minimal human capital loss - provides an exceptional lens for examining the economic repercussions of such events.

Central to this investigation is an exploration of the labor market implications of this disaster. Specifically, this research aims to answer how the labor market responds to drastic environmental or climate changes in the face of minimal human capital losses and how individuals adapt to such significant shifts in their environment. Because of the very limited physical destruction of property and human life, I can separate the heterogeneous nature of the disaster and understand how the labor market responds to such changes. 

This research is guided by two major schools of thought that appear to provide contrasting views. The first perspective draws on the principles of classical microeconomics, which conceptualizes the environment as a natural resource integrated into the production function as a form of capital. Consequently, negative shocks to such ``natural capital'' would be expected to drive wages and employment downward due to the quasi-complementarity of human and physical production components.

In contrast, the second perspective emerges from urban economics theory, particularly the Rosen-Roback model of urban spatial equilibrium. This model postulates that natural resources, such as a river, serve as amenities, enriching the quality of life and increasing utility generated by the inhabitants of the region. An exogenous negative shock to such an amenity, such as severe contamination, reduces individual utility levels. Consequently, firms may need to offset this decrease in utility by offering higher wages to retain their workers. Workers may consider relocating to regions offering better living conditions if the compensation is insufficient.

The Mariana disaster presents a unique opportunity to untangle these effects. The unique subtlety of the catastrophe lies in the toxic tailings' dual potential: on the one hand, they could impair the production function of industries relying on the river's resources, and on the other hand, they could pressure firms to increase wages as compensation for the depreciated amenity value of the region. The balance of these dynamics and their impacts on the labor market forms the core of this study.

This study employs a difference-in-differences model using a rich administrative panel dataset on the universe of formal workers in Brazil to quantify these effects. The approach compares municipalities directly impacted by the disaster with sufficiently similar yet unaffected ones. The primary focus is on the Mariana municipality itself as the treatment sample and the neighboring municipalities unaffected by the disaster serving as controls. The findings indicate a roughly 5.5 percent decrease in aggregate wages, a result that survives across all model variations. Interestingly, this wage reduction trend mirrors the heterogeneous effects observed in the agriculture and mining industries, which generally rely heavily on rivers and water, albeit at a lesser magnitude.

The observed effects extend to other regions significantly impacted by the disaster, namely municipalities through which the Doce River passes and municipalities in Espírito Santo's coastal area, although the impact varies in magnitude.

The study also investigates job retention and the probability of relocation by incorporating these aspects into a linear probability outcome in the DID model. The findings here offer limited insights, yet the overarching implications suggest that drastic environmental or climatic changes that do not directly threaten human life can still detrimentally impact the market's production function and reduce the population's overall wealth.

Nonetheless, the study also points out the presence of Rosen-Roback effects, that is, compensatory wage increases, particularly among worker groups whose roles are not directly dependent on water as a production component, such as those in office occupations. While these effects are indeed present among the potential market responses in the wake of the disaster, they appear neither strong nor representative enough to influence the aggregate level significantly.

My study contributes to the literature on the effects of disasters on labor market outcomes and urban spatial equilibrium. A  good example of a disaster being studied is Hurricane Katrina, that devasted New Orleans in 2005. \cite{groen_effect_2008} compares evacuees and individuals unaffected by the hurricane to find that initially, those affected suffered adverse labor market conditions but recovered quickly in the following months, especially if the individual decided to return to the region where the disaster occurred. Other studies also point out room for a limited recovery of some sectors in the affected regions and a ``bounce back'' behavior of labor market experiences of individuals \citep{vigdor_economic_2008, zissimopoulos_employment_2010}. \cite{mcintosh_measuring_2008} finds that the migration from New Orleans to Houston negatively affected the labor market outcomes of Houstonians. 

Other studies have explored the effects of droughts on the labor market, finding severe adverse effects on women's workdays and mobility \citep{efobi_long-term_2022, afridi_gendered_2022}.

On the other way, \cite{kirchberger_natural_2017} explores an earthquake that devasted the coastal region of Indonesia to find a resilient job market with positive wage growth driven by a limitation in labor supply in rural areas.

From the urban economics perspective, studies such as \cite{frame_housing_1998, ortega_rising_2018, boustan_effect_2020} explore variations in the environment, such as rising sea levels and flooding, to suggest the adverse effects of disasters on spatial equilibrium outcomes.

I fill the gap in the literature by exploring the interplay between capital shocks and spatial equilibrium dynamics, leveraging the unique nature of the Mariana Disaster. For instance, as a natural occurrence, the Katrina hurricane did not transform the environment to adverse potential toxic conditions. Moreover, the scale of the hurricane effectively destroyed a major city in the U.S., irrevocably mixing severe human capital shocks with changes in amenities. A recent accident that draws an interesting parallel to the Mariana Disaster is the East Palestine train derailment in Ohio, which in February 2023 released an enormous amount of toxic fumes into the atmosphere, particularly vinyl chloride, a substance supposedly carcinogenic to humans \citep{schnoke_economic_2023}. Moreover, to the best of my knowledge, this is the first study of the Mariana Disaster that employs rigorous econometric models to capture its causal effects on the labor market using heavily detailed data.

The remainder of the paper is organized as follows: Section \ref{sec:background} provides the background and reports on the government and other agents' reactions to the incident. Section \ref{sec:data} describes and explores the data used in my study. Section \ref{sec:identificationstrategy} is reserved for the identification strategy of the research and spatial data exploration. Section \ref{sec:empiricalstrategy} provides the empirical framework. Section \ref{sec:mainresults} and Section \ref{sec:discussion} present the main result together with its robustness checks and discusses the underlying mechanisms. Finally, Sector \ref{sec:conclusion} concludes. I reserve Appendix Section \ref{sec:theoryframe} to present the theoretical framework of Rosen-Roback spatial equilibrium and Factor of Productivity models.




\section{Background} \label{sec:background}

The Mariana disaster, which took place on November 5, 2015, is widely recognized as one of the most devastating environmental catastrophes in Brazilian and world history. Triggered by the collapse of the Fundão tailings dam, owned by Samarco (a joint venture between Vale S.A. and BHP Billiton), the disaster released an estimated 45 to 60 million cubic meters of iron ore extraction waste. Among other residues present, high mercury and other toxic heavy metal concentrations were found, generally used for mining operations \citep{hatje_environmental_2017}. The ensuing massive mudflow obliterated the small district of Bento Rodrigues, part of Mariana municipality in Minas Gerais.

The disaster's impacts extended far beyond the immediate vicinity of the dam. The mudflow traveled approximately 650 kilometers along the Doce River, reaching the Atlantic Ocean 17 days after the dam collapse \citep{hatje_environmental_2017}. This had significant implications for the coastal environment and marine life \citep{gabriel_contamination_2020}. In particular, the Espírito Santo region, known for its coastal fishery activities, was severely affected. The influx of iron ore waste led to a dramatic increase in water turbidity and heavy metal presence in marine life, which disrupted the photosynthesis process for aquatic plants and corals, ultimately reaching fish populations.

The Brazilian government's response to the disaster was multifaceted. Instants right before the disaster, it was issued an evacuation order for Bento Rodrigues' 600 inhabitants, effectively saving all but 19 lives. As the mud flowed, water distribution had to be interrupted, with the municipalities diverting water from other areas to supply their inhabitants. The government also issued a ban on fishing near the Doce River estuary and along the affected Atlantic Coast, potentially disrupting several markets.

Initially, Samarco, the company responsible for the incident, was fined approximately 66 million 2015 dollars. In March 2016, a settlement was reached in which Samarco, Vale, and BHP agreed to pay 5.3 billion dollars over 15 years to restore the environment and communities affected by the disaster completely. Recovery is estimated to take several decades \citep{fernandes_deep_2016}.

Despite its catastrophic scale, the disaster resulted in a surprisingly minimal loss of human life. The municipality of Mariana, which served as the epicenter of the disaster, is home to approximately 50,000 inhabitants. Remarkably, the majority of this population was spared, preserving a significant portion of the region's human capital. However, the environmental devastation inflicted upon the area was profoundly severe and, arguably, irrecoverable.

The disaster in Mariana provides a unique opportunity to investigate the interaction between a shift in the spatial equilibrium and labor market dynamics within the same market. Specifically, how external environmental shocks influence employment, wages, and worker mobility patterns. This insight is crucial for informing policymakers and preparing for future environmental disruptions.




\section{Data} \label{sec:data}

The data used for this study is the Annual Registry of Social Information (RAIS), an administrative panel data maintained by the Brazilian Ministry of Labor, containing information on the universe of the Brazilian formal labor market. I focus on the period from 2008 to 2018. Because the disaster occurred right at the end of the year of 2015, I consider, for the sake of simplicity, the year 2016 to be the first treatment year of my analysis.

The dataset provides a detailed description of its worker base. For instance, it identifies the individual through the Brazilian equivalent of a social security number, called PIS (Social Integration Program in Portuguese). Several labor market outcome determinants are present: gender, race, age, tenure in months, individual's education level, and nationality. Moreover, I observe the worker's occupation according to the Brazilian Occupation Code.

Other variables of interest are the type of work, which tells the employer-employee contract nature (if it is temporary or not). Workers in Brazil can have part-time or full-time jobs, expressed in the variable working hours per week. There is also the worker's hiring date, separation date, separation cause, and if the worker was present on December 31st of the observed year, the main variable that tells if a worker lost their job or not.

One of the key variables in my research is the employee's municipality of work, a strong indicator of their residency. The RAIS dataset also provides the worker's firm's identification code, CNPJ (National Registry of Legal Person, in Portuguese). CNPJ is strongly correlated with the municipality. A company with two branches in different locations appears as two separate CNPJs in the dataset.

Another crucial variable in RAIS is the economic activity code, CNAE (National Registry of Economic Activities, in Portuguese). CNAE code is a highly detailed categorization of the firm's main economic activity. Therefore, I can discriminate firms and groups of workers based on their ultimate production function's output. The code comprises seven numbers, each increasing a degree of detail. For instance, ``01'' represents agriculture activities, while ``0155505'' represents agriculture activities related to poultry, egg production.

\subsection{Generating Outcome Variables in RAIS}

The richness of RAIS allows for a comprehensive exploration of labor market dynamics in the aftermath of the disaster. To guide this investigation, I focus on three primary labor market outcome variables: average monthly wage, a ``mover'' indicator, and a dismissal indicator.

The core variable in this study, ``average monthly wage'', measures the typical monthly earnings a worker garners in a year. As is common with wage data, outliers may skew the analysis, and imputation errors may be present. To counteract these concerns, I narrow the scope to include only full-time workers working more than or equal to 20 hours weekly. Further, I apply a winsorization technique at the 2.5 and 97.5 percentiles to mitigate the impact of potential wage distortions. This method replaces the extremes of wage distribution with the nearest values within the defined percentiles, thereby reducing the influence of wage extremities or imputation errors on the final results.

Given the prevalence of multi-job holding among Brazilians, particularly those earning low or minimal wages, I ensure that the analysis pertains solely to primary jobs. I define the primary job as the one with the earliest hire date for an individual, as indicated in the dataset. In rare instances where multiple jobs share the same hire date, I default to the job with the highest pay. This data preparation procedure ultimately creates roughly individual-year-municipality spells.

The ``dismissal'' variable is a binary indicator tracking whether a worker maintained employment with the same firm until December 31st of a particular year. The variable assumes a value of one if the worker separates from the firm before the year-end; otherwise, it is zero.

The ``mover'' variable, developed by longitudinally tracking workers using their PIS, identifies individuals who change their work municipality across two consecutive years. Constructing this variable entails three steps. First, I identify individuals present in the research's region of interest in 2008. Then, I trace these individuals across all subsequent years, irrespective of their location in Brazil. This process repeats annually until 2018.

The second step involves examining individuals annually and noting any changes in their municipality of work. For each PIS, I compare the current municipality with the municipality in the following year, sorted by year and hire date. A change in municipality code leads to the assignment of a value of one to this new variable, while a lack of change results in zero.

The final step involves pruning observations not situated within the region of interest. As such, the effects quantified in this research should be understood as ``local market effects.'' This approach prioritizes the original location over individual identifiers, thereby ensuring that the analysis faithfully reflects the local labor market's dynamism.

My sample must also provide a basis for comparison across control and treatment groups. This implies ensuring a balance in all labor market outcome determinants employed in my study. In other words, I only consider instances where I observe at least one corresponding element in both cohorts. Such an approach is crucial for mitigating bias and ensuring the reliability of my analysis. I further elaborate on the nature of such determinants and how I deal with panel data imbalance in the sections dedicated to the identification and empirical strategy.




\section{Identification Strategy} \label{sec:identificationstrategy}

To estimate the causal effect of the Mariana disaster on labor market outcomes, I employ a difference-in-differences (DID) approach, which levers the variation in treatment status across time and between affected and non-affected municipalities. 

The main area of interest is the region called \emph{Quadrilátero Ferrífero} (roughly translated as Iron Square). The Mariana municipality, where the disaster took place, belongs to this region characterized by mining operations and its historical importance.

\begin{figure}[htb!]
    \centering
    \includegraphics[width=\textwidth]{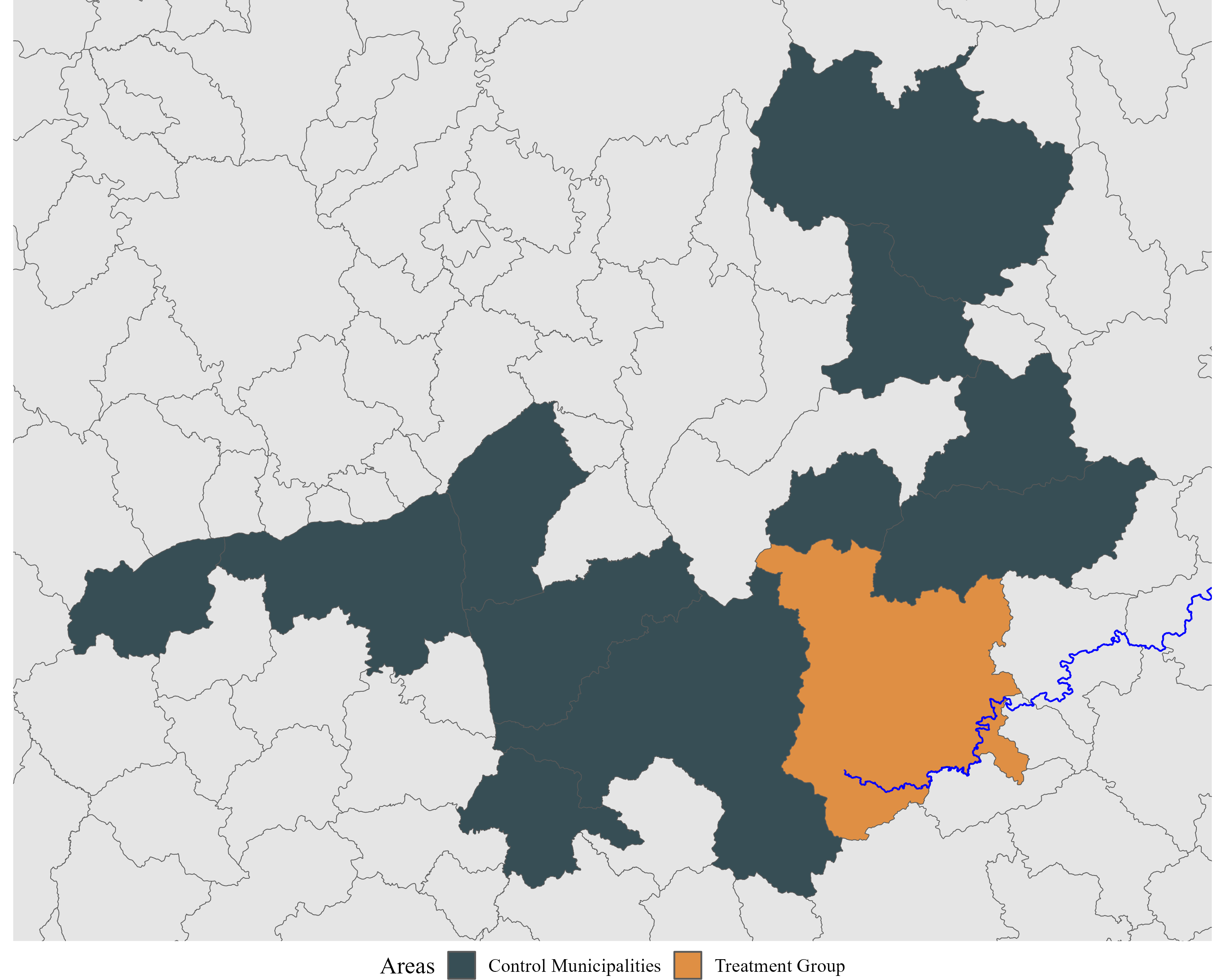}
    \caption{Overview of the Iron Square Region and the Doce River}
    \label{fig:fig_mariana_map}
\end{figure}

I exploit the fact that Doce River starts from Mariana and outflows the region, as shown in Figure \ref{fig:fig_mariana_map}. The brighter area corresponds to the Mariana Municipality, while the darker areas are the other municipalities belonging to the Quadrilátero Ferrífero region, which I use as the control group. The dark thicker line going outward from Mariana is the affected river. Given the event happened in November 2015, I consider 2015 the reference and 2016 until 2018 the post-treatment period. I also employ data from 2008 until 2014 as additional pre-treatment periods. It is possible to measure the causal effects using this method because the municipality where the individual is located is observed, together with the individual identifier. However, there are some concerns that arise with this identification strategy.

Although not as urban as the state capital metropolitan region, the region itself is sufficiently integrated to generate spillover effects due to individuals moving across municipalities. After the disaster, affected workers may have decided to move to locations and jobs that I use as the control, potentially distorting any measurement when employing the two groups. I address this issue with fixed effects interactions between social identifiers and the location of the individual. When spells related to the individual-municipality relationship are included in the model, the parameters absorb any effect related to geographical movement. Additional details are provided in Section \ref{sec:empiricalstrategy}.

Moreover, two effects may be at play when measuring the Mariana Disaster. The disaster itself, in other words, the dam rupture and the water contamination that eventually took over the Doce River. Separating these effects in this context is challenging, particularly when I ultimately intend to disentangle TFP and Rosen-Roback effects. I address this issue by employing the empirical analysis in two other regions affected by the disaster: the continuation of the Doce River until the Minas Gerais border and the river estuary region in the Espírito Santo state. Further explanation is provided in Section \ref{sec:discussion}.

Lastly, the dataset employed is highly unbalanced by nature, given that individuals may leave the labor market at any moment during the studied period, creating potential overlapping issues between treatment and control individuals. To counter this issue, I employ a doubly-robust regression where propensity score weights are used to balance both groups. I discuss this method in Section \ref{sec:empiricalstrategy}.

For a summary statistics of the region, refer to Appendix Section \ref{sec:sumstat}

\section{Empirical Strategy} \label{sec:empiricalstrategy}

The empirical strategy used in this paper exploits the quasi-random nature of the disaster to design a natural experiment setting. My identification strategy relies on the assumption that conditional on the control variables, the disaster is as good as randomly assigned across municipalities and individuals within the municipalities. I use a difference-in-differences approach to estimate the causal effect of the disaster, comparing the changes in labor market outcomes in the affected municipalities before and after the disaster to the changes in a control group of municipalities over the same period.

My main setting compares Mariana, the municipality where the disaster occurred, with municipalities from the same region of Minas Gerais, called the Iron Square Region\footnote{This is a rough translation from \emph{Quadrilátero Ferrífero} in Portuguese}. Later, I further explore the disaster by analyzing additional regions in the Discussion Section.

\subsection{Fixed Effects Models} \label{subsec:fixedeffects}

Here I present the specification for the fixed effects models. The basic version uses a two-way fixed effect regression using social identification and year fixed effects. I also present an alternative design of the baseline model to address potential spillover effects.

\subsubsection{Baseline Model}

The baseline specification is a fixed effects model that utilizes the individual social identification and year dummies. This model is outlined as follows:

\begin{equation}
    y_{imt} = \beta D_{imt} + \mu_i + \lambda_t + \epsilon_{imt}
\end{equation}

where $y_{imt}$ is the labor market outcome of individual $i$ at time $t$ and municipality $m$, $D_{imt}$ is a treatment indicator that equals 1 if individual $i$ is located in a disaster-affected municipality $m$ at time $t$ and 0 otherwise. The parameters $\mu_i$ and $\lambda_t$ capture time-invariant individual effects and yearly fixed effects, respectively. Lastly, $\epsilon_{imt}$ represents the error term that accounts for unobserved model characteristics. The key parameter of interest here is $\beta$, which quantifies the causal effect of the disaster.

The study covers the years from 2008 to 2018. As the disaster occurred in November 2015, and it took approximately one month for the waste to reach the ocean, 2016 is considered the first year of treatment for any region specification. The labor market outcomes for individual analysis are logarithmic wages and linear probabilities of being dismissed and moving from a municipality.

\subsubsection{Alternative Baseline Model: Interaction Model}

There are some potential challenges with the baseline model. Given the geographical proximity of the sampled municipalities and the intricate nature of the local economy, it's possible for individuals to relocate from the control to the treated region and vice versa. Such movement could introduce bias into the baseline model results. I propose an interaction model allowing individual interactions with their municipality to address this issue. This approach should capture the effects driven by changes in location, thereby allowing us to isolate the impact of the disaster on the local labor market. It is based on previous studies such as \cite{santanna_effects_2023, foged_immigrants_2016}. The new model is specified as follows:

\begin{equation}
    y_{imt} = \beta D_{imt} + \mu_i + \mu_m + \phi_{im} + \lambda_t + \epsilon_{imt}
\end{equation}

where the municipality fixed effects is $\mu_m$, similarly to the secondary model. The interaction term is $\phi_{im}$, where I create individual-municipality spells based on individual $i$ social identifier (pis) and the municipality geographical code from the Brazilian Institute of Geography and Statistics (IBGE).

\subsection{Augmented Inverse Propensity Score Weighting (Doubly Robust) Models}

I also implemented an augmented propensity score weighting procedure to increase the robustness of the fixed effects models. This procedure, also known as the Doubly-Robust approach, combines ordinary fixed effects regression with inverse propensity score weighting. This method addresses potential selection bias stemming from differences in the groups' support of covariates. The advantage of this method is the need for only one procedure to be correctly specified, the fixed-effects regression or the propensity score weighting, to prevent misspecification \citep{chernozhukov_doubledebiasedneyman_2017, robins_estimation_1994}.

My approach is similar to \cite{strittmatter_gender_2021}, adapted to a difference-in-differences design. The first step involves estimating the propensity score of each individual to belong to the treatment group. This is achieved using a logistic regression model with individual characteristics as the independent variables:

\begin{equation}
p(X_{i}) = Pr(D_i = 1 | X_i) = F(\theta'X_{i}) = \frac{e^{\theta'X_{i}}}{1 + e^{\theta'X_{i}}}
\end{equation}

where $X_i$ denotes a vector of time-invariant covariates for individual $i$ and $D_i = 1$ signifies that the individual belongs to the treatment group, namely the disaster-affected municipalities. The parameter $\theta$ establishes the logistic relationship between the covariates and which group the individual belongs to. In the next step, I predict $\hat{p(X_i)}$, the estimated propensity score, for all observations within the sample. The controls used are age, gender, race, education level, tenure, job occupation, and the firm's economic activity. Weights are then constructed by the following equation:

\begin{equation}
    \hat{W}_{it} = \frac{(1 - D_{it})\hat{p}(X_{i})}{1 - \hat{p}(X_{i})} \bigg/ \sum^{N}_{i=1} \frac{(1 - D_{it})\hat{p}(X_{i})}{1 - \hat{p}(X_{i})}
\end{equation}

Ideally, the weights assigned to individuals would remain unchanged across different years. However, given the potential for changes in the treatment and control sample over time, due to movement across municipalities and labor market leavers, I compute a set of weights for each year in the sample.

The final step minimizes the weighted sum of squares. Let $M = \{1, 2\}$ represent the two models specified in Subsection \ref{subsec:fixedeffects}, i.e., the baseline and the municipality-individual spell model, respectively. Let $\boldsymbol{\alpha_M}$ represent the corresponding fixed-effect set used for each specification. Therefore, the minimization problem is framed as follows:

\begin{equation}
\hat{\beta}_{dr}^M = \argmin_{\beta} \sum_{N} w_{it} (y_{imt} -\beta D_{imt} - f(\boldsymbol{\alpha_M}))^2
\end{equation}

where $w_{it} = \hat{W}_{it}$ is calculated using Equation 4. $N$ is the number of observations in the sample, $T$ is the number of time periods. The function $f(\boldsymbol{\alpha_M})$ represents a linear function wrapping the fixed-effects set:

\begin{align} 
    f(\boldsymbol{\alpha_1}) &= \alpha_i + \alpha_t \label{eq:set1} \\ 
    f(\boldsymbol{\alpha_2}) &= \alpha_i + \alpha_m + \alpha_{im} + \alpha_t \label{eq:set2}
\end{align}

As customary with difference-in-differences approaches \citep{mackinnon_cluster-robust_2023}, I cluster my standard errors at the municipality level, which is the dimension in the data that best represents the disaster's geographical level of impact.




\section{Main Results} \label{sec:mainresults}

In this section, I present the main results using the Mariana municipality as the treatment group, comparing similar municipalities in the ``Iron Square'' region. 

\vspace{10mm}

\begin{table}[htb!]
    \centering
        \caption{Main Results for Mariana Region Analysis}
        \label{tab:mainresults}
        \begin{adjustbox}{width=0.85\textwidth,center}
        \begin{tabular}[t]{lcccccc}
        \toprule
         & \multicolumn{2}{c}{Fixed Effects Models} & \multicolumn{2}{c}{Doubly Robust} \\
          & (1) & (2) & (3) & (4) \\
        \midrule
        Log Wage: Treat x Post & \num{-0.066}*** & \num{-0.052}*** & \num{-0.054}*** & \num{-0.055}***\\
         & (\num{0.016}) & (\num{0.010}) & (\num{0.013}) & (\num{0.009})\\
         \midrule
        Dismissed: Treat x Post & \num{-0.003} & \num{-0.005} & \num{0.041} & \num{0.010}\\
         & (\num{0.036}) & (\num{0.021}) & (\num{0.026}) & (\num{0.021})\\
        \midrule
        Individual FE & X & X & X & X\\
        Municipality FE &  & X &  & X\\
        Ind. x Municipality FE &  & X & & X\\
        Year FE & X & X & X & X\\
        N Clusters & \num{12} & \num{12} & \num{12} & \num{12}\\
        N & \num{1567721} & \num{1567721} & \num{1567721} & \num{1567721}\\
        \bottomrule
        \end{tabular}
        \end{adjustbox}
        \begin{flushleft}
            \parbox{\textwidth}{\footnotesize
            \textsuperscript{1} Standard-errors are clustered by municipality.\\
            \textsuperscript{2} * p < 0.1, ** p < 0.05, *** p < 0.01\\
            \textsuperscript{3} Covariates used for propensity score estimation in DR models are age, tenure, education level, gender, race, worker's occupation, and the firm's main economic activity.\\}
        \end{flushleft}
        
\end{table}

Table \ref{tab:mainresults} provides the regression results for the baseline and alternative baseline fixed effects models in Columns (1) and (2), respectively, and their Doubly Robust versions in Columns (3) and (4).

In terms of the Fixed Effects model, all specifications provide significant results at the 1\% level or lower for the effect of the disaster on wages (Log Wage: Treat x Post). The coefficient estimate values at -0.066 for the baseline in Column (1) and -0.052 when adding the interactive individual-municipality fixed effects parameter in Column (2), suggesting a notable decrease in wages after the disaster in the treated areas.

The results' significance and direction are insensitive to the DR regression set, with -0.054 in Column (3), representing the doubly robust version of the baseline model, and -0.055 for the DR version of the interacted fixed effects model, all at the 1\% significance level.

Even though the log-wage study provided significant evidence of negative effects, the effect on dismissals (Dismissed: Treat x Post), however,  appears insignificant in all model specifications, with most magnitudes being negligible except for the doubly robust baseline in Column (3), indicating that the disaster did not significantly alter the dismissal rate in the treated municipality of Mariana, at least when comparing to the neighboring, non-affected, municipalities.

These results provide strong initial evidence of a substantial negative impact on wages due to the Mariana Dam disaster in its own municipality. One could be inclined to affirm that, given the negative nature of wage effects, it was, in the first place, a capital shock in accordance with the Factor of Production hypothesis. Nevertheless, these results are not sufficient to disentangle the Factor of Production and Rosen-Roback effects. There is concern about the disaster taking place in the region, creating massive destruction of capital when the dam was destroyed, and potentially devastating mining operations or economic activities surrounding the incident area. Therefore, these results capture not only the total contamination of the river but also the disaster itself. 

\subsection{Robustness Checks}

Before proceeding with the mechanisms and heterogeneous effects underlying the results, I discuss the survivability of my procedure when applying typical robustness checks used in difference-in-differences designs. In this section, I investigate whether the results yielded by the previous log-wage regressions are due to randomness in the control municipalities. 

\subsubsection{Event Studies}

The event study methodology serves to validate the common trends assumption, a fundamental prerequisite for employing the Difference-in-Differences (DID) identification strategy \citep{callaway_difference--differences_2021-1, goodman-bacon_difference--differences_2021}. In essence, this assumption states that, in the hypothetical scenario where no treatment was administered, both the control and treated groups would exhibit parallel trends over time. However, the challenge arises when examining post-treatment periods, as the counterfactual outcome is unobservable. Nevertheless, I test the pre-treatment trends to make sure the null hypothesis, that the control and treated groups exhibit the same estimate trend, is not rejected.

Transforming the event study method into a regression model allows us to dissect the treatment effect parameter across distinct time periods. This strategy requires the identification of a reference period, ideally just before treatment, to compare the dynamic estimates. In this study, I choose 2015 as the pre-disaster reference year, primarily due to the timing of the event. The disaster unfolded in November, taking less than a month to reach the ocean, which suggests that the bulk of 2015 remained unaffected by the dam rupture. Consequently, 2015 is regarded as the reference year.

If the impact took some time to manifest in the labor market, we would expect it to become apparent in the subsequent year, affirming my specification's validity. On the other hand, if the shock was immediate, my choice could be seen as a 'conservative' approach. This method would likely result in estimates biased towards zero, as we are potentially incorporating part of the post-treatment period into our pre-treatment reference period. 

The empirical specification is as follows:

\begin{align}
y_{imt} &= \sum_{{t=2008}}^{2018} \beta_t D_{imt} + f(\boldsymbol{\alpha_M}) \quad \text{for} \quad t \neq 2015
\end{align}

where $\beta_t$ is the decomposed estimate parameter of interest. $f(\alpha_M)$ represents a linear function with the corresponding fixed-effects model from equations \ref{eq:set1} and \ref{eq:set2}. I also present event study versions for the doubly robust alongside the two baseline models, further reinforcing my robustness checks.

\begin{figure}
    \centering
    \caption{Event Studies for Log Wage Regression Models}
    \label{fig:es_logwage_ironbelt}
    \begin{subfigure}[b]{0.45\textwidth}
        \includegraphics[width=\textwidth]{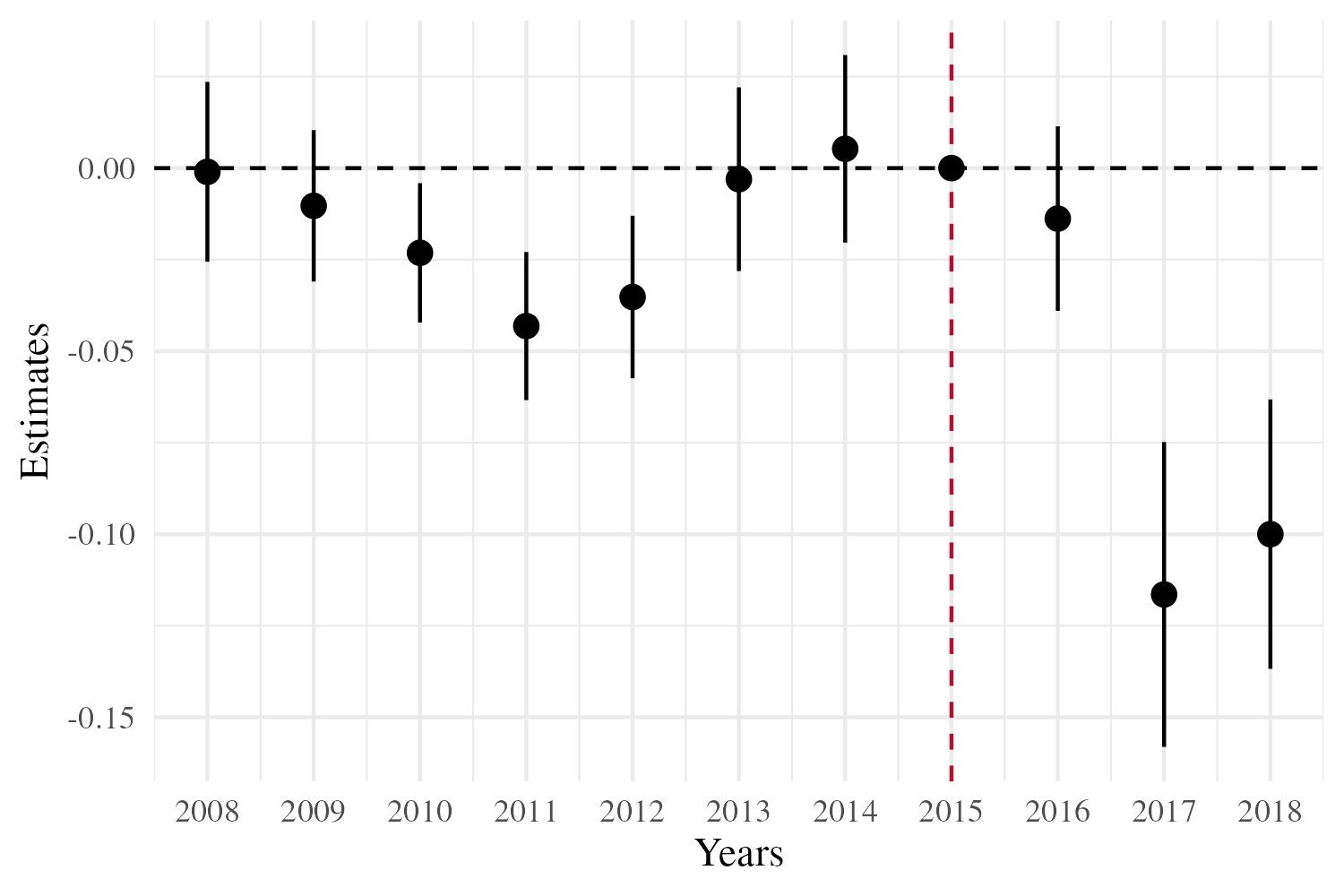}
        \caption{Individual Fixed Effect ES}
        \label{fig:ironbelt__log_wage_pis}
    \end{subfigure}
    \hfill
    \begin{subfigure}[b]{0.45\textwidth}
        \includegraphics[width=\textwidth]{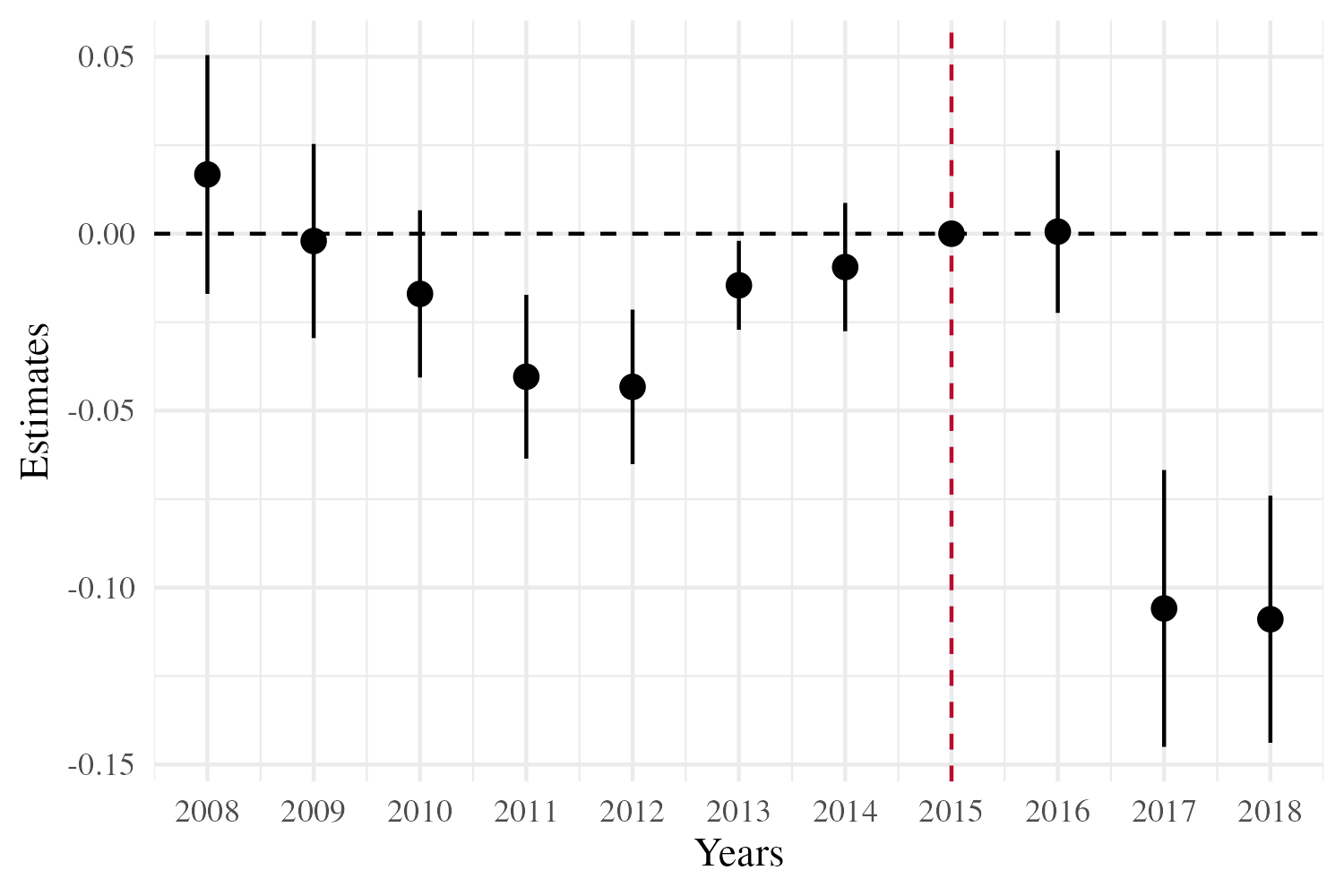}
        \caption{Individual Fixed Effect ES - DR}
        \label{fig:ironbelt__log_wage_ipw_pis}
    \end{subfigure}
    
    \begin{subfigure}[b]{0.45\textwidth}
        \includegraphics[width=\textwidth]{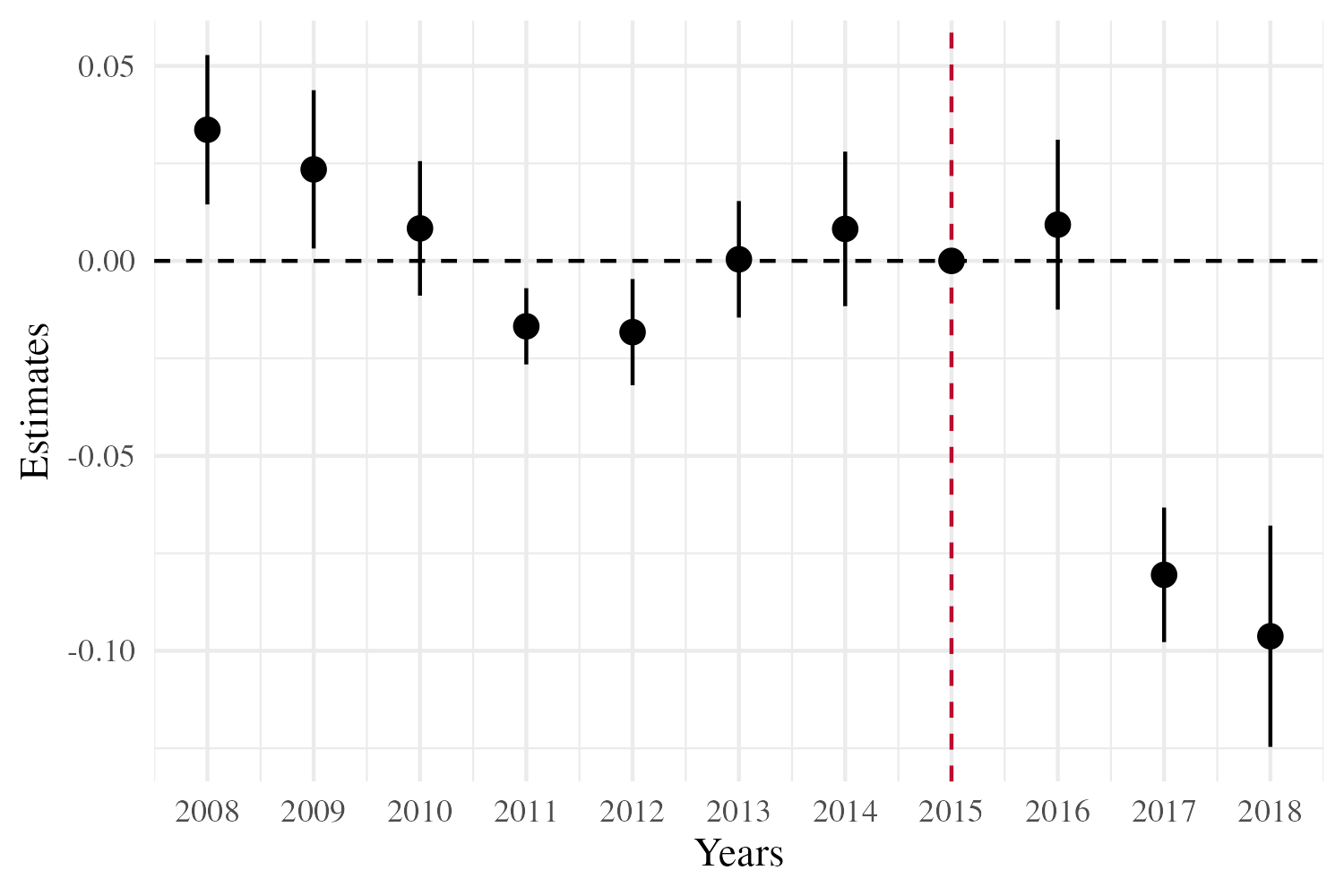}
        \caption{Ind. x Municipality Fixed Effect ES}
        \label{fig:5}
    \end{subfigure}
    \hfill
    \begin{subfigure}[b]{0.45\textwidth}
        \includegraphics[width=\textwidth]{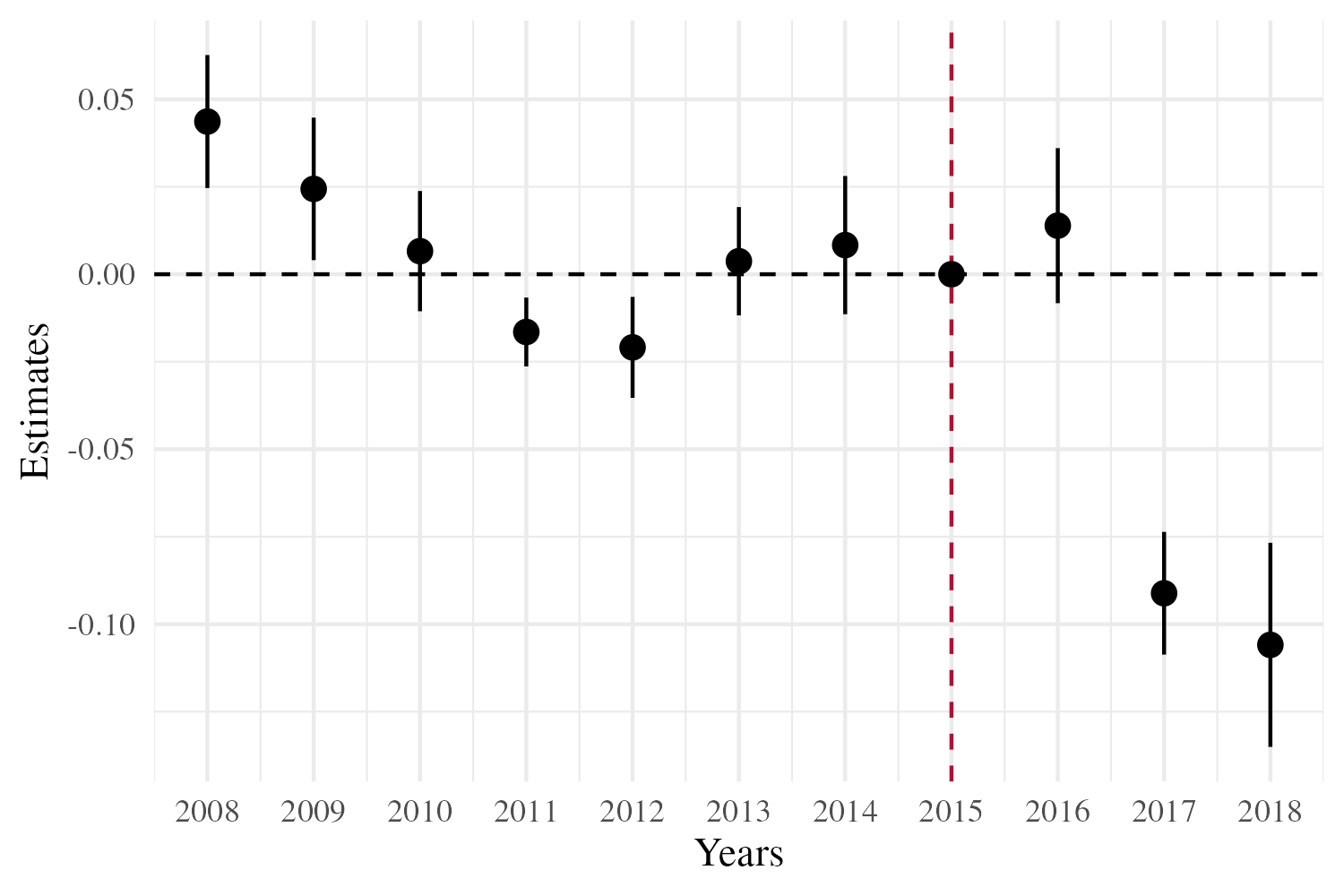}
        \caption{Ind. x Municipality Fixed Effect ES - DR}
        \label{fig:6}
    \end{subfigure}
\end{figure}

Figure \ref{fig:es_logwage_ironbelt} presents the plot panel of all models. The first column represents the baseline version, while the second column shows regressions from the doubly robust specification. The first row is the simplest model where only individual fixed effects are included to control for time-invariant characteristics. The second row represents the model with the individual and municipality interaction terms.

In general, all results are robust with different degrees of success. The trends do not change dramatically with different settings. Notably, in any model, there is no difference between the reference year and the estimate in 2016. There are possible explanations behind this phenomenon. The immediate effect could have taken some time to appear in the aggregate market. There could also be conflicting heterogeneous effects between two industries that would absorb the shock differently, such as healthcare and the mining industry. Nevertheless, the result that appears in the following years indicates the lasting effects in the aggregate market are related to capital shocks, in accordance with the Factor of Production hypothesis.

\subsubsection{Placebo Tests}

The effects observed in the regressions could be driven by random variations in the control group. Given Mariana is the sole treated region in the main setting, with another eleven municipalities serving as control, it is a natural assumption that perhaps the observed estimates are generated by some municipalities having disturbances after the disaster.

\begin{figure}
    \centering
    \includegraphics[width=0.7\textwidth]{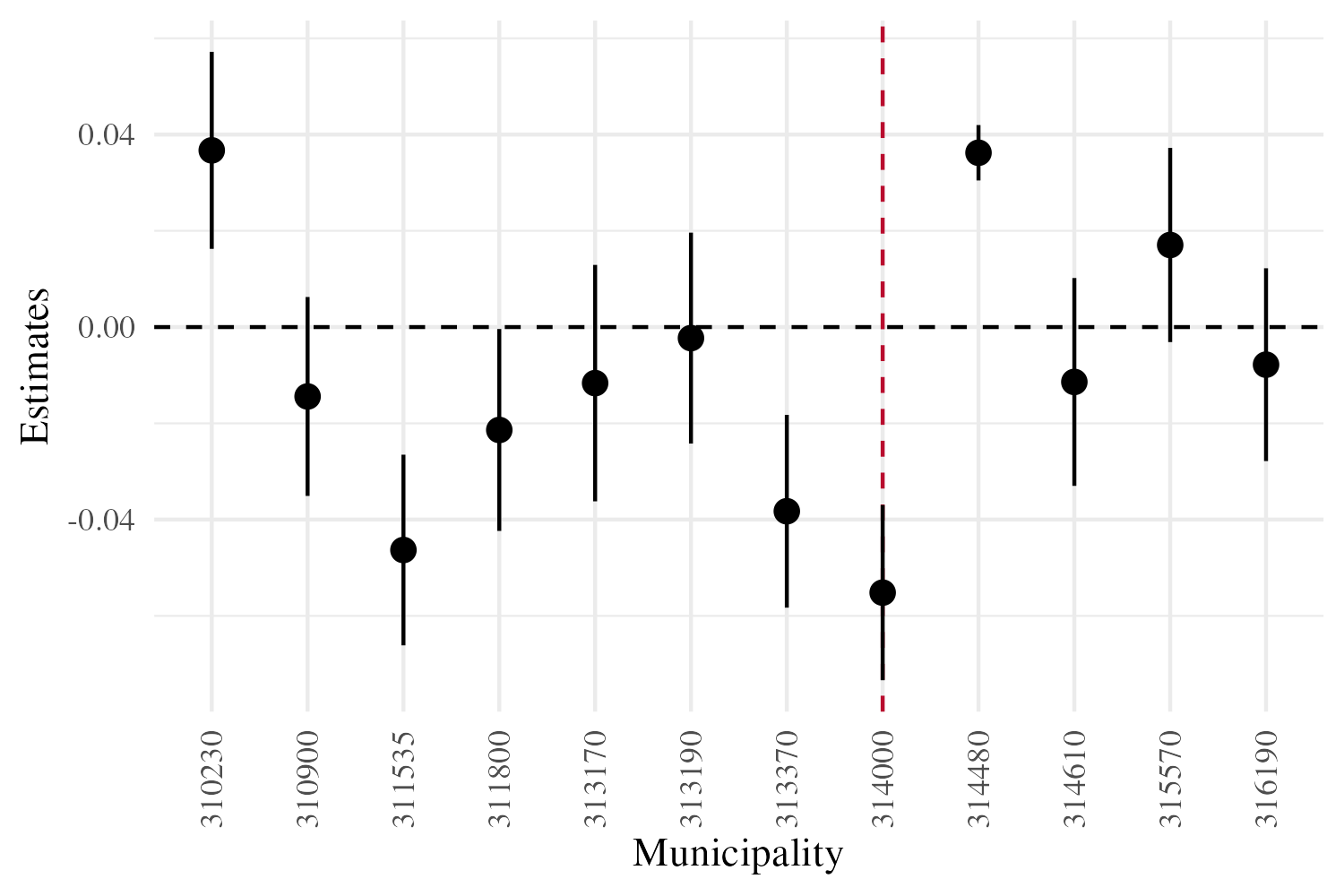}
    \caption{Permutation Tests for Ind. x Municipality FE in a Doubly Robust Design}
    \label{fig:placebo_wage}
\end{figure}

To address this issue, I elaborate a permutation test where I eliminate Mariana from the sample and permute each municipality as if they were treated instead. This is similar to the robustness approaches proposed in synthetic control methods, such as \cite{abadie_comparative_2015, abadie_synthetic_2010, abadie_economic_2003}. Figure \ref{fig:placebo_wage} shows the placebo test result.

The horizontal axis represents the municipality code used as the placebo treatment for my most ``robust'' model, corresponding to the DR design with municipality-individual spells in the fixed effects parameters. The dashed vertical line represents Mariana's estimate, the original treated group. When performing the placebo tests, I remove Mariana from the sample and test the selected placebo with the remaining municipalities. Ideally, all other municipalities' results would be centered at zero, with Mariana being by far the most affected and isolated. However, the result reveals that seven out of eleven municipalities reject the null hypothesis. Even though it has some degree of undesirability, looking carefully at the estimates, Mariana still has the largest magnitude, in accordance with the fact it is where the disaster occurred. Moreover, the placebos are, in the majority, negatively biased, meaning that my original main result is potentially biased toward zero.

The most problematic municipalities are 310230 and 314480, Alvinópolis, and Nova Lima, with positive estimates that could negatively drive the main results even further. However, the two municipalities do not have sufficient representatives in the sample to drastically change the effect. 

 The placebo analysis allows the conclusion that a considerable proportion of the effects measured by the main regressions are driven by the Mariana disaster, not some randomness happening in the control municipalities.




\section{Discussion} \label{sec:discussion}

This section discusses the mechanisms underpinning the principal findings of this study. Initially, I explore the Factor of Productivity hypothesis through a heterogenous effect approach, specifically focusing on industries strongly connected to water usage and dam operations. Moreover, I also test the healthcare labor market, as it serves as an illustrative case to ascertain a positive labor demand alteration after the accident that is unrelated to Rosen-Roback effects. There is some evidence that individuals related to clinics and hospitals registered a positive change in wages, presumably due to the surge in labor demand triggered by health-related contingencies in the local populace.

The subsequent segment investigates the influence on the extended region. Using spatial data obtained from the Institute of Brazilian Geography and Statistics, the state of Minas Gerais is partitioned into its respective water basins. This allows for a comparative analysis between municipalities in the Doce River water basin directly impacted by the catastrophe and those in unrelated water basins, with particular care taken to exclude Mariana or the municipalities in the Iron Square from the sample. The main objective of this secondary study is to disentangle the effects of the dam rupture itself from the consequences of water pollution. This distinction serves to highlight the dynamics between Factor Productivity and Rosen-Roback models in the context of severe environmental perturbations.

In the final portion of this section, I focus on ``pure Rosen-Roback effects'' observed in certain occupations and economic activities. While these effects are discernible at the micro level, there is no evidence of it at the aggregate market level, aligning with the insights thus far learned from previous results, where individuals, on the aggregate market, absorb negative shocks in wages when facing a dismal change in environment.

\subsection{Verifying the Factor of Production Hypothesis}

The Factor of Production Hypothesis states that water is a component of the production function of Mariana and, therefore, when the dam rupture contaminated the river, it disrupted several economic activities in the region that use water in their framework. The main industries with potential affected outcomes are mining, agriculture, forestry, and fishing.

To measure the heterogeneous effect of these industries, I interact the treatment indicator function with another dummy variable that values one when the individual works in the aforementioned industry. I show the modified baseline models in the following equation:

\begin{align}
y_{imt} &= \beta D_{imt} \times I_{if} + f(\alpha_M) + \epsilon_{imt}
\end{align}

where $I_{if}$ is the new indicator function for when firm $f$ of worker $i$ belongs to the industry of interest. I use the economic activity code variable in RAIS to identify iron mining, agriculture, forestry, and fishing. The latter three activities, however, are not sufficient for testing on their own, and therefore I aggregate them into ``rural activities''. Iron mining operations correspond to the 07 code, while agriculture, forestry, and fishing correspond to 01, 02, and 03, respectively.

I also provide an additional test where I measure the impact of the disaster on healthcare workers. This is another robustness check to see if the results follow the main intuition. Even though the disaster was minimal in direct human capital loss, the environmental destruction and water contamination are catalysts for negative mental health and well-being shocks. Therefore, healthcare workers are expected to experience an increase in wages or negative effects on dismissals, driven by the sharp increase in the market's labor demand. The corresponding code for healthcare firms is 86, 87, and 88.

\subsubsection{Water as capital: Rural and Mining Activities}

Table \ref{tab:heterogeneous} presents the results of the heterogeneous effects by economic activities, particularly focusing on iron mining, rural activities (agriculture, forestry, and fishing), and the healthcare industry, which are likely the most affected by the event. I provide the fixed-effect baseline models in the first two columns, while in the last batch, I incorporate propensity score weights, as before, for the doubly robust framework. Each column represents a different fixed-effect approach in the same fashion as the main results model. Each row represents a heterogeneous effect based on the specified industries.

\begin{table}[htb!]
    \vspace{10mm}
    \centering
        \caption{Heterogeneous Effect Analysis by Economic Activities for Mariana Region}
        \label{tab:heterogeneous}
        \begin{adjustbox}{width=0.85\textwidth,center}
            \begin{tabular}[t]{lcccccc}
            \toprule
            & \multicolumn{4}{c}{Log Wage}\\
            \midrule
             & \multicolumn{2}{c}{Fixed Effects Models} & \multicolumn{2}{c}{Doubly Robust Models} \\
              & (1) & (2) & (3) & (4)\\
            \midrule
            Iron Mining: Treat x Post & \num{-0.196}*** & \num{-0.297}*** & \num{-0.240}*** & \num{-0.294}***\\
             & (\num{0.017}) & (\num{0.009}) & (\num{0.028}) & (\num{0.005})\\
            \midrule
            Rural Activities: Treat x Post & \num{-0.063}** &  \num{-0.055}*** & \num{-0.025} & \num{-0.025}\\
             & (\num{0.023})  & (\num{0.011}) & (\num{0.023}) & (\num{0.020})\\
            \midrule
            Healthcare: Treat x Post & \num{0.080}*** & \num{0.074}*** & \num{0.090}*** & \num{0.085}***\\
             & (\num{0.015})  & (\num{0.012}) & (\num{0.019})  & (\num{0.023})\\
            \midrule
            & \multicolumn{4}{c}{Linear Probability of Moving}\\
            \midrule
            Iron Mining: Treat x Post & \num{0.081} & \num{0.075}** & \num{0.141}*** & \num{0.122}***\\
             & (\num{0.052}) & (\num{0.024}) & (\num{0.017}) & (\num{0.021})\\
            \midrule
            Rural Activities: Treat x Post & \num{-0.011} & \num{-0.028} & \num{0.018} & \num{-0.001}\\
             & (\num{0.031}) & (\num{0.022}) & (\num{0.014}) & (\num{0.016})\\
            \midrule
            Healthcare: Treat x Post & \num{-0.082}* & \num{-0.086}*** & \num{-0.046}*** & \num{-0.064}***\\
             & (\num{0.041}) & (\num{0.022}) & (\num{0.013}) & (\num{0.015})\\
            \midrule
            Individual FE & X & X & X & X\\
            Individual x Municipality FE &  & X &  & X\\
            Municipality FE &  & X &  & X\\
            Year FE & X & X & X & X\\
            N Clusters & \num{12} & \num{12} & \num{12} & \num{12}\\
            N & \num{1567721} & \num{1567721} & \num{1567721} & \num{1567721}\\
            N (Mover) & \num{1348847} & \num{1348847} & \num{1348847} & \num{1348847} \\
            \bottomrule
            \end{tabular}
        \end{adjustbox}
        \begin{flushleft}
            \parbox{\textwidth}{\footnotesize
            \textsuperscript{1} Standard-errors are clustered by municipality.\\
            \textsuperscript{2} * p < 0.1, ** p < 0.05, *** p < 0.01\\
            \textsuperscript{3} Covariates used for propensity score estimation in DR models are age, tenure, education level, gender, race, worker's occupation, and the firm's main economic activity.\\
            \textsuperscript{4} Iron Mining activity code is 07. Rural Activities correspond to agriculture (01), forestry (02), and fishing (03). Healthcare code is 86, 87, and 88.\\}
        \end{flushleft}
\end{table}

Iron mining, being the primary industry directly linked to the disaster, shows a strong, negative wage effect across all models. Notably, the baseline models in Columns (1) and (2) reveal a negative, statistically significant effect of the treatment after the disaster, ranging from -0.196 to -0.297. These figures indicate a substantial reduction in wages in the post-disaster period, potentially reflecting the direct impacts of the disaster on this industry, including the suspension of operations, workforce reduction, lowered productivity, and the loss of the dam itself.

This observation is mirrored in the doubly robust models in Columns (3) and (4). In this case, the estimated wage effects are slightly more pronounced, ranging from -0.240 (Column 3) to -0.294 (Column 4). In other words, there was at least a 20 percent decrease in wages in the mining industry in the disaster aftermath, with a considerable amount of this magnitude directly related to it.

In contrast, rural activities exhibit another, albeit similar, pattern. While the wage effect remains negative, the magnitude is markedly smaller compared to that of the iron mining industry. The fixed effects models suggest an effect ranging from -0.063 (Column 1) to -0.055 (Column 2), while the doubly robust models give an estimate of -0.025 (Columns 4 and 6). However, the DR specification does not yield statistical significance. Given that the accident happened in a small specific region (the district of Bento Rodrigues), and I measured the entire municipality, not only the mentioned district, these results are the first evidence of a negative capital shock related to the pollution, not the dam destruction. Still, the significance level reveals that the results yielded may be driven mainly by outcome determinants imbalance between control and treatment groups.

\subsubsection{The disaster as a catalyst for labor demand: The healthcare service}

Table \ref{tab:heterogeneous} also provides results related to healthcare sector wage variations, showing a significant increase in wages post-treatment periods across both the fixed effects and doubly robust models. This is likely due to the surge in demand for healthcare services following the disaster, requiring a bolstering of the healthcare workforce in the region and consequently leading to wage increases.

Nevertheless, the possibility exists that these positive outcomes are reflections of "Rosen-Roback" effects, propelled by compensatory wage adjustments in response to shocks to local amenities. This hypothesis can be examined more closely by assessing the probability of migration. By tracking the same individual across time periods or even within the same time period but in distinct jobs, we can determine if their subsequent location differs from their current one. If a change in location is identified, a value of one is assigned to this dummy variable; if not, it is assigned a zero. This approach provides insight into the mobility patterns of individuals in the wake of such environmental disasters and helps discern if the wage increases are indeed an outcome of Rosen-Roback effects.

The regression models are the same as those used for log-wage outcomes. The outcome of interest, in this case, is the dummy variable representing whether the individual moved in a linear probability model fashion. Results are presented in the second set of regressions in Table \ref{tab:heterogeneous}. Results for the mining industry suggest the probability of an individual leaving Mariana after the disaster increased from 7.5 to 14.1 percent, in accordance with the economic intuition of individuals leaving due to severe disruption in production. The precision of the results increases when employing the doubly robust procedure. For agriculture, estimates did not yield any significant measurement across all models.

For the healthcare industry, all results are negative, with varying degrees of significance, revealing that the probability of moving for a healthcare worker decreased after the disaster. The negative direction of these estimates shows the ambiguity of the effects. It could be influenced by a surge in labor demand due to the health shock caused by the disaster or as a consequence of the compensatory wages at play.

\subsection{Extended Regions Analysis}

The Mariana region sample inherently intertwines the immediate disaster and its ensuing environmental pollution. As such, the results derived so far can be regarded as a composite of capital destruction and subsequent environmental fallout at best. To scrutinize human behavior in the face of disruptive environmental changes with minimal direct human capital loss, I turn my attention to regions solely affected by the pollution of the Doce River and the Atlantic Ocean. If these regions exhibit findings akin to those of the Mariana region, it will serve as further evidence supporting the proposition that the overall behavior of the Factor of Productivity hypothesis tends to eclipse Rosen-Roback effects in such circumstances. Figure \label{fig:extended_regions} shows the new regions explored.

The empirical strategy applied is the same as before, fixed effect models augmented with propensity score weights. However, due to the different geographical locations, I propose an alternative version of the identification strategy for the two additional samples, albeit still maintaining the difference-in-differences approach.

\subsubsection{Identification Strategy for the extended region within Minas Gerais}

The first extended region under study encompasses municipalities situated along the Doce River but external to the Quadrilátero Ferrífero region. This is visually depicted in Figure \ref{fig:extended}, wherein the brighter highlighted municipalities represent the treatment group adjacent to the polluted river. 

I exploit the fact that the Doce River serves as the primary water body of its water basin to select the control group. Using spatial data from the Minas Gerais government for the municipality and water basin borders, I am able to identify municipalities with their water system unrelated to the disaster area. Specifically, I choose the Paranaiba River and Grande River water basins. They are situated diametrically opposite the Doce River and are not directly connected, ensuring that there is a sufficient sample of individuals comparable to the treatment group and a similar climatic backdrop conducive to analogous economic activities in both groups, but also making sure these municipalities will not suffer from spillover effects due to the Mariana Disaster.

\subsubsection{Identification Strategy for the Espírito Santo State}

The disaster's final region of impact was Espírito Santo state, the convergence point of the Doce River and the Atlantic. By December 2015, pollution had navigated the river's course, culminating in the contamination of the ocean. This event resulted in a government-imposed prohibition on local fishing activities, a considerable repercussion given the substantial number of coastal fishermen in the area.

\begin{figure}[htp]
    \centering
    \begin{subfigure}[b]{\textwidth}
        \centering
        \includegraphics[width=\textwidth, keepaspectratio, height=0.4\textheight]{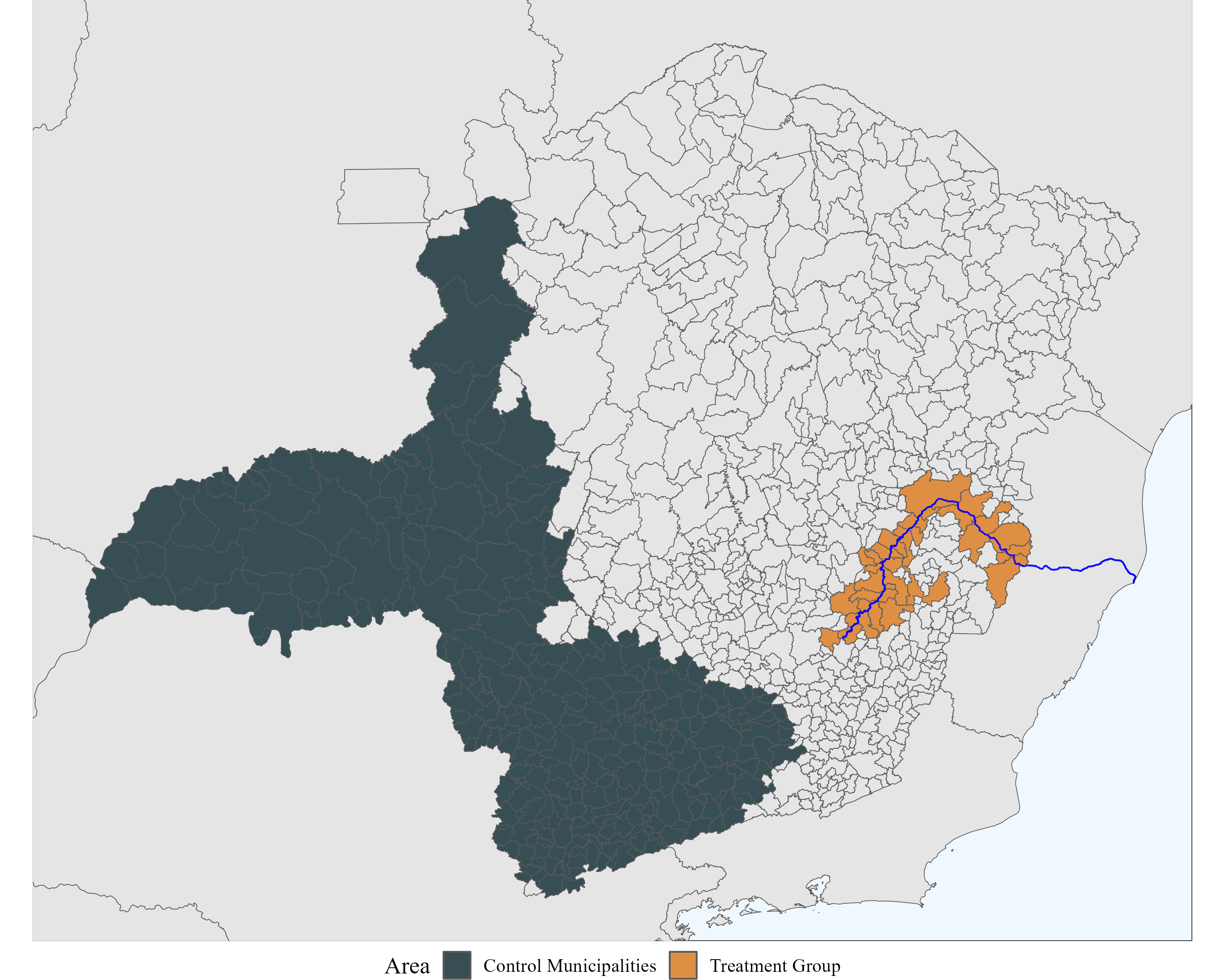}
        \caption{Extended Region of Doce River in Minas Gerais.}
        \label{fig:extended}
    \end{subfigure}

    \vspace{5mm} 

    \begin{subfigure}[b]{\textwidth}
        \centering
        \includegraphics[width=\textwidth, keepaspectratio, height=0.4\textheight]{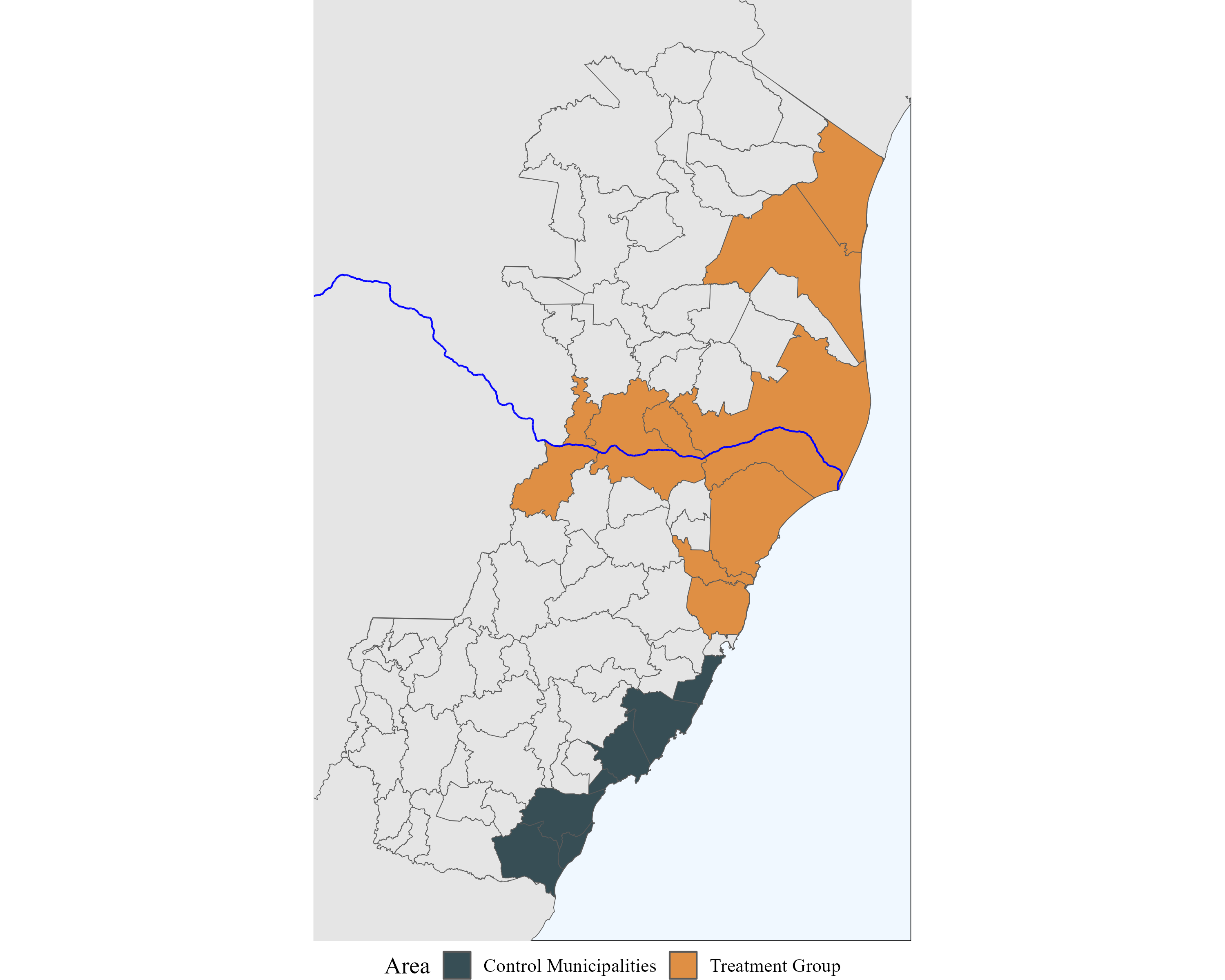}
        \caption{Espírito Santo State and Affected Regions Near the Atlantic Ocean.}
        \label{fig:es}
    \end{subfigure}

    \caption{Additional regions analyzed. The darkened areas are control municipalities.}
    \label{fig:extended_regions}
\end{figure}

Based on the government's disaster impact reports, which state the contaminated ocean water went northwards towards Bahia, my analysis distinguishes this region into a northern and southern coast. Consequently, the treated units comprise individuals in municipalities crossed by the river or those residing in the northern coastal municipalities identified by the government as having contaminated seawater, as shown in Figure \ref{fig:es}.

The control group consists of the southern coastal municipalities unaffected by the pollution and located south of the state capital, Vitória.

My ultimate objective is to extract the effect of the disaster in these regions without the unintended effects of capital disruption of the Iron Square sample. Summary statistics tables for both settings are displayed in the Appendix Section.

\subsubsection{Extended Regions Results}

Table \ref{tab:extendedlogwage} presents the results using log-wage as the outcome variable. The first row shows the results for the aggregate market in the extended region of Minas Gerais. The four columns stand for the same models used in the previous analyses. In the extended Minas Gerais sample, I find that the Mariana disaster significantly negatively impacted wages, as indicated across all models, ranging from approximately 3.8 to 5.8 percent. This trend is followed, to some degree, by the rural activities sector, shown in the second row. However, the effects lose significance when accounting for the balance between control and treatment groups, which means most of these effects, at least for this specific sector, appear due to a lack of counterfactual among the two tested groups. Still, the sign of all estimates is negative, in a similar magnitude range from the aggregate market.

Even though the rural market shows weak evidence of an effect, the same directions of both aggregate and rural markets suggest the environment, represented here by Doce River, plays much more the role of a productivity component in the labor market than an amenity, per Mariana's main results. 

Moving onto the Espírito Santo sample, shown in the second part of Table \ref{tab:extendedlogwage}, while the aggregate effect on log wages presented in the third estimate row is negligible for all models, the sector-specific analysis reveals a different picture. The rural activities in the treated municipalities from Espírito Santo suffered significant negative wage effects across all models, with at least ten percent of statistical significance. When I focus on fishing activity firms only, as shown in the sixth estimate row, the magnitude increases, ranging from around 6.6 and 7.0 percent decrease for the basic fixed effects models to 11.3 and 12.5 for the doubly robust models. Not surprisingly, these wage results suggest that after the river and now the ocean were spoiled, the market's production function had a negative capital shock, pushing wages downwards.

\begin{table}[htb!]
    \centering
        \caption{Extended Regions Results}
        \label{tab:extendedlogwage}
        \begin{adjustbox}{width=0.85\textwidth,center}
            \begin{tabular}[t]{lcccc}
                    \toprule
                    & \multicolumn{4}{c}{Log Wage}\\
                    \midrule
                    & \multicolumn{2}{c}{Fixed Effects Models} & \multicolumn{2}{c}{Doubly Robust Models} \\
                    & (1) & (2) & (3) & (4)\\
                    \midrule
                     \textbf{Extended Minas Gerais Sample}  & & & &\\
                    \midrule
                    Aggregate: Treat x Post & \num{-0.038}*** & \num{-0.035}** & \num{-0.058}*** & \num{-0.056}**\\
                    & (\num{0.014}) & (\num{0.015}) & (\num{0.021}) & (\num{0.023})\\
                    \midrule 
                    Rural Activities: Treat x Post & \num{-0.061}*** & \num{-0.045}*** & \num{-0.047}* & \num{-0.038}\\
                    & (\num{0.012}) & (\num{0.010}) & (\num{0.027}) & (\num{0.029})\\
                    \midrule
                    N Clusters & \num{323} & \num{323} & \num{323} & \num{323}\\
                    N & \num{19334590} & \num{19334590} & \num{19334590} & \num{19334590}\\
                    \midrule
                    \textbf{Espírito Santo Sample}  & & & &\\
                    \midrule
                    Aggregate: Treat x Post & \num{0.013} & \num{0.003} & \num{0.015} & \num{0.003}\\
                    & (\num{0.009}) & (\num{0.012}) & (\num{0.013}) & (\num{0.017})\\
                    \midrule
                    Rural Activities: Treat x Post & \num{-0.035}*** & \num{-0.027}** & \num{-0.033}*** & \num{-0.024}*\\
                    & (\num{0.008}) & (\num{0.010}) & (\num{0.010}) & (\num{0.012})\\
                    \midrule
                    Fishing: Treat x Post & \num{-0.066}*** & \num{-0.070}*** & \num{-0.113}*** & \num{-0.125}***\\
                    & (\num{0.010}) & (\num{0.009}) & (\num{0.019}) & (\num{0.016})\\
                    \midrule
                    N Clusters & \num{16} & \num{16} & \num{16} & \num{16}\\
                    N & \num{5100644} & \num{5100644} & \num{5100644} & \num{5100644}\\
                    \bottomrule
                    Individual FE & X & X & X & X\\
                    Municipality FE &  & X &  & X\\
                    Individual x Municipality FE &  & X &  & X\\
                    Year FE & X & X & X & X\\
                    \bottomrule
                    \multicolumn{5}{l}{\rule{0pt}{1em}\textsuperscript{1} Standard-errors are clustered by municipality.}\\
                    \multicolumn{5}{l}{\rule{0pt}{1em}\textsuperscript{2} * p < 0.1, ** p < 0.05, *** p < 0.01}\\
                    \bottomrule
                \end{tabular}
        \end{adjustbox}
        \begin{flushleft}
            \parbox{\textwidth}{\footnotesize
            \textsuperscript{1} Standard-errors are clustered by municipality.\\
            \textsuperscript{2} * p < 0.1, ** p < 0.05, *** p < 0.01\\
            \textsuperscript{3} Covariates used for propensity score estimation in DR models are age, tenure, education level, gender, race, worker's occupation, and the firm's main economic activity.\\
            \textsuperscript{4} Rural Activities correspond to agriculture (01), forestry (02), and fishing (03).\\}
        \end{flushleft}
\end{table}

I also further investigate the extended region by providing estimates when using the dismissal and the moving variable in Table \ref{tab:extendedmovers}. The first four columns (1-4) are for the linear probability of being dismissed from one's current job, while the last set (5-8) shows the linear probability of moving out from a municipality results. Accordingly, I explore the two additional regions, the extended Minas Gerais and the Espírito Santo coastal region. 

For dismissals, the vast majority of the results are insignificant, except for the extended Minas Gerais aggregate market, where it shows a positive change in dismissal probability ranging from 1.9 to 2.3 percent. Even though the magnitude and direction of the estimates can be explained by the productivity shock hypothesis, they do not survive my most robust model of fixed effect interaction and inverse propensity score weighting in Column (4).

In a similar fashion, the rural activity analysis for the extended Minas area yielded a negative 7.9 percent chance, statistically significant at 5 percent, when using a doubly robust procedure. However, the effect disappears when accounting for interactions between individual and municipality fixed effects. For Espírito Santo, no estimates yielded statistically significant results for the aggregate market and for rural activities. For rural activities, results are negligibly close to zero in magnitude.

Casting our attention to the probability of moving analysis for the extended Minas area, results are inconclusive, with the aggregate market and rural activities both displaying meaningful effects in Column (7) but not surviving interactions between individuals and municipalities, which suggests the calculated causal effects are driven by changes in the treatment and control composition when individuals move from one place to another.

In contrast, in the Espírito Santo sample, I find a positive effect on moving out probabilities, even with inconclusive results in the dismissal analysis. One explanation could be that the moving-out probability appears conditional on already dismissed individuals. In other words, a worker's position in the fishing firm may not be affected by the disaster, only through wages. However, in the case of a dismissal, this individual is more likely to find another job in a firm outside of the original municipality.

The adverse effects on job security and population stability appear to be less meaningful than the direct productivity losses from capital destruction. Still, it points towards a critical interaction between the environment (as a form of capital component) and local economic conditions. In other words, in a disaster like Mariana's, where the environment is severely spoiled, but human capital is preserved, which is a reflection of drastic environmental changes potentially already occurring due to massive human intervention everywhere, it should be expected an overall impoverishment of the market instead of a search for reallocation.

\begin{sidewaystable}[h!]
    \centering
        \caption{Extended Regions Results - Probability of Dismissals and Moving Out}
        \label{tab:extendedmovers}
        \begin{adjustbox}{width=0.85\textwidth,center}
            \begin{tabular}[t]{lcccccccc}
                    \toprule
                    & \multicolumn{4}{c}{Prob. of Dismissal} & \multicolumn{4}{c}{Prob. of Moving}\\
                    \midrule
                    & \multicolumn{2}{c}{Fixed Effects Models} & \multicolumn{2}{c}{Doubly Robust Models} & \multicolumn{2}{c}{Fixed Effects Models} & \multicolumn{2}{c}{Doubly Robust Models} \\
                    & (1) & (2) & (3) & (4) & (5) & (6) & (7) & (8)\\
                    \midrule
                     \textbf{Extended Minas Gerais Sample}  & & & &\\
                    \midrule
                    Aggregate:Treat x Post & \num{0.019}** & \num{0.023}** & \num{0.019}** & \num{0.020} & \num{0.011}* & \num{0.010} & \num{0.015}*** & \num{0.008}\\
                    & (\num{0.008}) & (\num{0.009}) & (\num{0.009}) & (\num{0.012}) & (\num{0.006}) & (\num{0.007}) & (\num{0.005}) & (\num{0.006})\\
                    \midrule 
                    Rural Activities: Treat x Post & \num{-0.011} & \num{0.027} & \num{-0.079}** & \num{-0.004} & \num{-0.026}  & \num{0.009} & \num{-0.055}* & \num{0.010}\\
                    & (\num{0.029}) & (\num{0.023}) & (\num{0.040}) & (\num{0.054}) & (\num{0.019}) & (\num{0.014}) & (\num{0.028}) & (\num{0.031})\\
                    \midrule
                    N Clusters & \num{323} & \num{323} & \num{323} & \num{323} & \num{323} & \num{323} & \num{323} & \num{323}\\
                    N & \num{19334590} & \num{19334590} & \num{19334590} & \num{19334590}  & \num{16541725} & \num{16541725} & \num{16541725} & \num{16541725}\\
                    \midrule
                    \textbf{Espírito Santo Sample}  & & & &\\
                    \midrule
                    Aggregate: Treat x Post & \num{-0.021} & \num{-0.023} & \num{-0.035} & \num{-0.024} & \num{-0.018}& \num{-0.016} & \num{-0.035} & \num{-0.017}\\
                     & (\num{0.031}) & (\num{0.015}) & (\num{0.026}) & (\num{0.016}) & (\num{0.040}) & (\num{0.010}) & (\num{0.035}) & (\num{0.012})\\
                    \midrule
                    Rural Activities: Treat x Post & \num{0.008} & \num{0.010} & \num{0.006} & \num{0.009} & \num{-0.004} & \num{0.009} & \num{-0.005} & \num{0.009}\\
                     & (\num{0.024}) & (\num{0.028}) & (\num{0.024}) & (\num{0.028}) & (\num{0.020}) & (\num{0.026}) & (\num{0.020}) & (\num{0.026})\\
                    \midrule
                    Fishing: Treat x Post & \num{0.006} & \num{0.021} & \num{-0.105} & \num{-0.074} & \num{0.076}** & \num{0.094}*** & \num{0.049} & \num{0.101}**\\
                    & (\num{0.029}) & (\num{0.033}) & (\num{0.063}) & (\num{0.083})  & (\num{0.034}) & (\num{0.017}) & (\num{0.057}) & (\num{0.036})\\
                    \midrule
                    N Clusters & \num{16} & \num{16} & \num{16} & \num{16} & \num{16} & \num{16} & \num{16} & \num{16}\\
                    N & \num{5100644} & \num{5100644} & \num{5100644} & \num{5100644} & \num{4351754} & \num{4351754} & \num{4351754} & \num{4351754}\\
                    \bottomrule
                    Individual FE & X & X & X & X & X & X & X & X\\
                    Municipality FE &  & X &  & X &  & X &  & X\\
                    Individual x Municipality FE &  & X &  & X &  & X &  & X\\
                    Year FE & X & X & X & X & X & X & X & X\\
                    \bottomrule
                    \multicolumn{5}{l}{\rule{0pt}{1em}\textsuperscript{1} Standard-errors are clustered by municipality.}\\
                    \multicolumn{5}{l}{\rule{0pt}{1em}\textsuperscript{2} * p < 0.1, ** p < 0.05, *** p < 0.01}\\
                    \bottomrule
                \end{tabular}
        \end{adjustbox}
        \begin{flushleft}
            \parbox{\textwidth}{\footnotesize
            \textsuperscript{1} Standard-errors are clustered by municipality.\\
            \textsuperscript{2} * p < 0.1, ** p < 0.05, *** p < 0.01\\
            \textsuperscript{3} Covariates used for propensity score estimation in DR models are age, tenure, education level, gender, race, worker's occupation, and the firm's main economic activity.\\
            \textsuperscript{4} Rural Activities correspond to agriculture (01), forestry (02), and fishing (03).\\}
        \end{flushleft}
\end{sidewaystable}

Nevertheless, given the size and complexity of the studied markets, there are potentially ignored underlying effects when observing specific occupations or using other dimensions in the regressions. Even though, on the aggregate, it seems individuals absorb the shock through wages, there is still the possibility of ``Rosen-Roback'' effects when accounting for specific demographics or types of firms.

\clearpage

\subsection{Office Occupations Analysis}

Even though the measurements indicate that the effects surfacing on the aggregate market are related to capital shocks, there is a possibility that other heterogeneous effects are at play, which I did not capture in my previous regressions. To understand better the Mariana disaster's implications on the equilibrium relationship between wages and amenities, I perform an occupation-based analysis of workers in the sampled areas. 

In accordance with the theoretical framework shown in Section \ref{sec:theoryframe}, I choose occupations as the key variable for the analysis due to, in this case, the individual making the decision to stay or move from the affected regions. The challenge of using a firm's economic activities is that, within the same firm, there will be too heterogeneous cohorts of workers at different education levels, types of occupations, and tenure duration. Therefore, a more correct approach is to perform the study taking into account the worker's occupation types across firms.

It is possible to identify the type of occupation through RAIS. I use Brazil's Ministry of Labor's Brazilian Classification of Occupations, CBO in Portuguese, to classify the worker-wise type of activity. The code can be granulated up until six digits, each subsequent digit representing a more specific set of tasks. For instance, the code ``21'' represents STEM professionals, while ``214'' represents engineers and architects. A final specification example for this code sequence is ``2142-10'', representing airport engineers.

Potentially, there are two main types of occupations. Occupations directly and indirectly related to water as a component of labor. For example, farmers primarily rely on water to perform their activities and ultimately provide labor output. On the other hand, office workers, such as accountants and clerks, may only take the water contamination as an amenity nuisance if, for the sake of simplification, we ignore the costs of acquiring freshwater.

This class of occupations \footnote{The codes I use for the regressions are 410, 411, 412, 413, 414, and 415. These represent clerks, secretaries, and office administrators, generally responsible for the day-to-day routine operations of offices but not necessarily high-specialized functions such as accountants, lawyers, or bankers.}, which is less directly linked to the local environment in its day-to-day operations compared to the rural activities and fishing sectors, offers a suitable case study of potential Rosen-Roback effects.

Office occupations are primarily influenced by factors such as technology availability, human capital, and infrastructure rather than local natural resources. Therefore, we would typically expect this sector to be less sensitive to environmental disruptions. However, if the Rosen-Roback framework holds, we would anticipate some level of impact from the Mariana disaster, primarily through indirect channels such as reduced local consumption or changes in local labor supply dynamics. Moreover, firms may increase their wages specifically for these occupations to compensate for the negative shock on surrounding amenities, which is the key measurement of my study.

My empirical framework, structured similarly to previous sections, utilizes occupation code dummies as interaction terms with the treatment variable. The four columns in Table \ref{tab:office} represent the two baseline models and two doubly robust models. Each pair includes an individual fixed effect term and an interaction term between municipality and individual identifiers.

The analysis is carried out across the three key regions, each yielding distinctive insights about the labor market outcomes post-disaster.

The first region of focus is the Iron Square region, where proper Mariana is located and where the disaster took place. Here, the data indicate a shift in spatial equilibrium, as office workers' wages show an increase of approximately 5 percent in the more robust models. Further evidence supporting this equilibrium shift is the 3-6 percent decrease in the likelihood of office workers moving or being dismissed from their jobs when compared to unaffected municipalities in the region.

The subsequent region, the extended region of Minas, displays distinct patterns. While wage change remains neutral, we observe a one percent increase in the likelihood of dismissal. This trend is consistent across all model specifications and aligns with a similar pattern in the probability of moving.

Lastly, the Espírito Santo state analysis offers weak evidence with a slight two percent wage increase at the ten percent significance level. The changes in the likelihood of dismissal and migration out of the region are negligible and statistically insignificant when compared with the control groups.

Despite the variations across regions, these results provide compelling evidence that spatial equilibrium dynamics significantly shape labor market outcomes after environmental shocks, at least for specific categories indirectly related to the environment per se. However, it is crucial to understand these findings in context. As established in the main results, the aggregate market tends to respond to such shocks as capital shocks above all else.

\begin{table}[h!]
    \centering
        \caption{Office Occupation Analysis}
        \label{tab:office}
        \begin{adjustbox}{width=0.85\textwidth,center}
\begin{tabular}[t]{lcccccc}
\toprule
\toprule
& \multicolumn{2}{c}{Fixed Effects Models} & \multicolumn{2}{c}{DR Models}\\
\midrule
  & (1) & (2) & (3) & (4)\\
\midrule
\textbf{Mariana Municipality Region} & & & & \\
\midrule
Log Wage: Treat x Post & \num{0.014} & \num{0.020} & \num{0.052}** & \num{0.051}**\\
 & (\num{0.018}) & (\num{0.011}) & (\num{0.023}) & (\num{0.023})\\
\midrule
Dismissal: Treat x Post & \num{-0.042} & \num{-0.048}** & \num{-0.037}** & \num{-0.054}***\\
 & (\num{0.029})  & (\num{0.018}) & (\num{0.014})  & (\num{0.010})\\
\midrule
Mover: Treat x Post & \num{-0.059} & \num{-0.062}** & \num{-0.033}** & \num{-0.052}***\\
 & (\num{0.042}) & (\num{0.021}) & (\num{0.013}) & (\num{0.014})\\
\midrule
N Clusters & \num{12} & \num{12} & \num{12} & \num{12}\\
N & \num{1567721} & \num{1567721} & \num{1567721} & \num{1567721}\\
N (mover) & \num{1348847} & \num{1348847} & \num{1348847} & \num{1348847}\\
\midrule
\textbf{Extended Minas Gerais Region} & & & & \\
\midrule
Log Wage: Treat x Post & \num{0.001}  & \num{0.004} & \num{0.008}  & \num{0.013}\\
 & (\num{0.011})  & (\num{0.010}) & (\num{0.010})  & (\num{0.009})\\
\midrule
Dismissal: Treat x Post & \num{0.012}***  & \num{0.012} & \num{0.012}***  & \num{0.011}\\
 & (\num{0.003})  & (\num{0.008}) & (\num{0.003})  & (\num{0.008})\\
\midrule
Mover: Treat x Post & \num{0.014}***  & \num{0.012}** & \num{0.012}***  & \num{0.011}**\\
 & (\num{0.004})  & (\num{0.006}) & (\num{0.003})  & (\num{0.005})\\
\midrule
N Clusters & \num{323}  & \num{323} & \num{323} & \num{323}\\
N & \num{19334590} & \num{19334590} & \num{19334590} & \num{19334590}\\
N (mover) & \num{16541725} & \num{16541725} & \num{16541725} & \num{16541725}\\
\midrule
\textbf{Espírito Santo Region} & & & & \\
\midrule
Log Wage: Treat x Post & \num{0.023}* & \num{0.022}* & \num{0.025}* & \num{0.022}\\
 & (\num{0.012}) & (\num{0.012}) & (\num{0.013}) & (\num{0.014})\\
\midrule
Dismissal: Log Wage: Treat x Post & \num{-0.012} & \num{-0.017}* & \num{-0.021} & \num{-0.018}\\
 & (\num{0.016}) & (\num{0.009}) & (\num{0.023}) & (\num{0.012})\\
\midrule
Mover: Treat x Post & \num{0.007} & \num{0.002} & \num{-0.004} &  \num{0.000}\\
 & (\num{0.017}) & (\num{0.006}) & (\num{0.026}) & (\num{0.007})\\
\midrule
N Clusters & \num{16} & \num{16} & \num{16} & \num{16}\\
N  & \num{5100644} & \num{5100644} & \num{5100644} & \num{5100644}\\
N (mover) & \num{4351754} & \num{4351754} & \num{4351754} & \num{4351754}\\
\midrule
Individual FE & X & X & X & X\\
Individual x Municipality FE &  & X &  & X\\
Municipality FE &  & X & & X\\
Year FE & X & X & X & X\\
\bottomrule
    \end{tabular}
        \end{adjustbox}
        \begin{flushleft}
            \parbox{\textwidth}{\footnotesize
            \textsuperscript{1} Standard-errors are clustered by municipality.\\
            \textsuperscript{2} * p < 0.1, ** p < 0.05, *** p < 0.01\\
            \textsuperscript{3} Covariates used for propensity score estimation in DR models are age, tenure, education level, gender, race, worker's occupation, and the firm's main economic activity.\\
            \textsuperscript{4} The codes I use for the regressions are 410, 411, 412, 413, 414, and 415.\\}
        \end{flushleft}
\end{table}



\clearpage

\section{Conclusion} \label{sec:conclusion}

This study offers a comprehensive analysis of the Mariana disaster's impact on labor market outcomes, juxtaposing two theoretical models in economics: the productivity factor model and the Rosen-Roback spatial equilibrium model. The former posits that the disaster, through water contamination, negatively shocks physical capital levels, thereby driving down wage levels. Conversely, the latter model suggests that the river, as an amenity, should prompt an increase in wages to maintain workers' utility levels in the wake of its destruction; otherwise, workers may relocate to more desirable areas.

Employing a rich administrative dataset encompassing all formal workers and utilizing a difference-in-differences design, the findings indicate significant negative effects on wages in two out of the three studied regions. This suggests that the river primarily functions as capital in the market's aggregate production function. These results persist across alternative fixed effect specifications and ``doubly'' robust regressions using inverse propensity score weights for estimate calculation.

The results of examining heterogeneous effects across specific industries align with economic intuition. For sectors such as mining, agriculture, and fisheries, where water is integral to operations, the impact on workers' wages was also negative, with varying degrees of robustness. This indicates that the aggregate results mirror underlying disruptive capital shocks closely tied to these economic activities. However, the linear probability models for job movement and dismissal in the aggregate and aforementioned specific markets did not yield strong enough results to establish a conclusive scenario.

Interestingly, when focusing on individuals working in occupations not directly related to water, specifically office workers such as clerks and secretaries, positive pressure on wages and a higher propensity to leave the affected area were observed, aligning with the urban spatial equilibrium proposal.

These findings suggest that, on aggregate, individuals experiencing drastic environmental changes with negligible alterations in human capital, such as potential scenarios of climate change effects, may absorb the shock through the market's production function, leading to wage losses and overall impoverishment. However, the effects of spatial equilibrium should not be underestimated. Policymakers must consider a given region's workforce composition and key characteristics when formulating responses to such environmental disasters.

There is considerable room for more studies regarding the Mariana Disaster. Due to its massive disruption and the complexity of the region, the labor market itself represents one of many aspects potentially affected by the incident. For instance, when analyzing spatial equilibrium, one should also consider the influence of housing costs and the real estate market. However, I was unable to acquire related data.

In conclusion, this study provides an effort to understand the relationship between environmental disasters, labor market outcomes, and spatial equilibria, highlighting the need for further research to disentangle these effects and inform policy decisions. As climate change continues to pose significant threats to our environment, understanding these dynamics will be crucial for mitigating the economic impacts and fostering resilience in affected communities.

\clearpage
\setlength\bibsep{0pt}
\bibliographystyle{econ}
\small \bibliography{references}

\clearpage

\doublespacing
\appendix
\renewcommand\thesection{A.\arabic{section}}
\section*{Appendix}
\section{Theoretical Framework} \label{sec:theoryframe}

The core of this analysis rests on  two established frameworks of economics, namely the productivity framework and the Rosen-Roback model. These two perspectives allow us to explore the multi-faceted impact of the Mariana disaster, from productivity shocks to spatial equilibrium, and provide us with an understanding of the economic implications of the event. Through these lenses, I can illustrate how similar environmental catastrophes affect the economy through individual firms' decision-making and individual decision-making processes.

\subsection{Productivity Framework} \label{fpframework}

The productivity framework forms the foundation of our understanding of a firm's decision-making process, which in turn impacts overall economic output. A widely utilized tool in this context is the Cobb-Douglas production function, which allows us to model the relationship between capital, labor, and output. In its most basic form, a firm's optimization problem can be expressed as follows:

\begin{align}
\max_{K,L} \quad Y - r K &- w L \
& \text{s.t.} \quad Y = A K^\alpha L^{1-\alpha}
\end{align}

Here, $Y$ represents total output, $A$ is the Total Factor of Production (TFP), $K$ stands for the capital stock, $L$ denotes the labor force, and $\alpha$ is the output elasticity of capital. The price of labor, or wages ($w$), is determined by the firm's optimization of resource allocation. By solving the maximization problem, we get the following First Order Condition:

\begin{equation}
w = (1 - \alpha) A \left( \frac{K}{L} \right)^\alpha
\end{equation}

From this, we can observe that the wage level is directly proportional to the level of capital, given $\alpha > 0$. Within the scope of this research, water is treated as a component of productivity, in other words, capital. Consequently, the exogenous shock caused by the dam rupture, which resulted in a decrease in available fresh water and related resources, is expected to exert downward pressure on wages.

\subsection{Rosen-Roback Model} \label{subsec:theoryrosenroback}

The Rosen-Roback model is instrumental in understanding the spatial equilibrium across regions or cities, achieved when workers and firms are indifferent between locations. This balance comes from the interplay between wages, rents, amenities, and transportation costs. Specifically, individuals maximize their utility derived from these factors when considering different locations. For example, an individual $i$ will prefer to stay in location $a$ rather than moving to location $b$ if:

\begin{equation}
U_i(W_a) + U_i(A_a) - U_i(R_a) > U_i(W_b) + U_i(A_b) - U_i(R_b) - U(T_{ab})
\end{equation}

In this equation, $W_a$, $R_a$, and $A_a$ represent the wage, rental cost, and perceived level of amenities, respectively, at location $a$, and similarly, $W_b$, $R_b$, and $A_b$ for location $b$. $T_{ab}$ stands for the transportation cost of moving from $a$ to $b$, and $U_i$ is a quasi-concave and monotonic utility function. This model predicts that a decrease in perceived amenities could lead individuals to bear transportation costs to move away, prompting firms and landlords to react by increasing wages or decreasing rents, respectively, to retain the population and achieve a new equilibrium.

In the aftermath of the Mariana disaster, it can be hypothesized that office workers and other individuals who do not rely on water for their productivity but perceive the river as an amenity might choose to relocate unless compensated through increased wages or decreased housing costs.

By applying these two theoretical frameworks, I can capture both the direct and indirect impacts of the Mariana disaster on local productivity and spatial equilibrium, providing a comprehensive picture of the economic consequences of such environmental shocks.

\clearpage

\section{Samples' Summary Statistics Tables} \label{sec:sumstat}

Tables \ref{tab:marianastat}, \ref{tab:minasstat}, and \ref{tab:esstat} present summary statistics for the three samples used in this study: the Mariana region sample, known as the Quadrilátero Ferrífero, the extended Minas Gerais sample, which is the Water Basin's analysis sample, and the Espírito Santo sample, used for measuring the impacts on the coastal region. These statistics provide a snapshot of the demographic and economic characteristics of the control and treatment groups in each sample in 2015, right before the disaster.

Table \ref{tab:marianastat} provides summary statistics for the Mariana region sample. The mean age and wage in the treatment group were slightly lower than in the control group. The treatment group also had a higher percentage of Black, Other, and Pardo individuals and a lower percentage of White individuals. The gender distribution was more balanced in the treatment group, and the education level was slightly lower. Notably, it shows the presence of mining operations in the area, with at least 12 percent of the labor market being composed of mining firms.

Table \ref{tab:minasstat} presents summary statistics for the extended Minas Gerais sample. The mean age was slightly lower in the treatment group, while the mean wage was slightly higher. The treatment group had a higher percentage of Pardo individuals and a lower percentage of White individuals. The gender distribution was almost identical in the control and treatment groups. 

Table \ref{tab:esstat} provides summary statistics for the Espírito Santo sample. The mean age and wage were slightly higher in the treatment group. The treatment group also had a higher percentage of Black, Other, and Pardo individuals, and a lower percentage of White individuals. The gender distribution was more skewed towards males in the treatment group, and the education level was slightly lower.

In both the extended Minas Gerais and Espírito Santo samples, the data reveals that mining operations are virtually non-existent. This observation is crucial for disentangling the effects of the dam rupture from the impacts of the ensuing contamination. The evidence strongly suggests that the negative effects observed in the extended Minas Gerais sample are attributable to contamination rather than disruption of mining operations.

\begin{table}[h!]
    \centering
        \caption{Mariana Region Sample Summary Statistics}
        \label{tab:marianastat}
        \begin{adjustbox}{width=\textwidth,center}
\begin{tabular}[t]{lllrr}
\toprule
  &    &     & Control  & Treatment \\
\midrule
 & Age & Mean & \num{34.80} & \num{34.52}\\
 & Wage & Mean & \num{2135.30} & \num{1845.28}\\
   \midrule
Race & Black & Percent & \num{7.80} & \num{10.44}\\
 & Other & Percent & \num{10.11} & \num{14.48}\\
 & Pardo & Percent & \num{46.05} & \num{44.81}\\
 & White & Percent & \num{36.05} & \num{30.27}\\
   \midrule
Gender & Female & Percent & \num{33.97} & \num{38.81}\\
 & Male & Percent & \num{66.03} & \num{61.19}\\
   \midrule
Education & College Education & Percent & \num{9.62} & \num{7.96}\\
 & High School Education & Percent & \num{57.55} & \num{54.40}\\
 & No High School Diploma & Percent & \num{32.83} & \num{37.65}\\
   \midrule
Industry & Construction & Percent & \num{18.95} & \num{13.14}\\
 & Financial Services & Percent & \num{0.94} & \num{0.82}\\
 & Mining & Percent & \num{17.59} & \num{12.33}\\
 & Public Administration & Percent & \num{1.38} & \num{0.91}\\
 & Rural Activities & Percent & \num{1.46} & \num{1.93}\\
 & Tourism & Percent & \num{5.79} & \num{6.76}\\
  \midrule
 & All & N & \num{146719} & \num{12389}\\
\bottomrule
\end{tabular}
\end{adjustbox}
\end{table}

\begin{table}[h!]
    \centering
        \caption{Extended Minas Gerais Sample Summary Statistics}
        \label{tab:minasstat}
        \begin{adjustbox}{width=\textwidth,center}
\begin{tabular}[t]{lllrr}
\toprule
  &    &     & Control & Treatment\\
\midrule
 & Age & Mean & \num{35.13} & \num{34.83}\\
 & Wage & Mean & \num{1518.98} & \num{1523.94}\\
 \midrule
Race & Black & Percent & \num{7.02} & \num{5.97}\\
 & Other & Percent & \num{8.20} & \num{6.10}\\
 & Pardo & Percent & \num{20.58} & \num{40.59}\\
 & White & Percent & \num{64.20} & \num{47.35}\\
 \midrule
Gender & Female & Percent & \num{37.78} & \num{37.67}\\
 & Male & Percent & \num{62.22} & \num{62.33}\\
 \midrule
Education & College Education & Percent & \num{7.55} & \num{7.12}\\
 & High School Education & Percent & \num{47.48} & \num{57.96}\\
 & No High School Diploma & Percent & \num{44.97} & \num{34.93}\\
 \midrule
Industry & Construction & Percent & \num{7.09} & \num{10.32}\\
 & Financial Services & Percent & \num{1.27} & \num{1.54}\\
 & Mining & Percent & \num{0.24} & \num{0.00}\\
 & Public Administration & Percent & \num{0.74} & \num{0.03}\\
 & Rural Activities & Percent & \num{16.17} & \num{4.35}\\
 & Tourism & Percent & \num{4.79} & \num{5.42}\\
 \midrule
 & All & N & \num{1586451} & \num{235242}\\
\bottomrule
\end{tabular}
\end{adjustbox}
\end{table}

\begin{table}[h!]
    \centering
        \caption{Espírito Santo Sample Summary Statistics}
        \label{tab:esstat}
        \begin{adjustbox}{width=\textwidth,center}
\begin{tabular}[t]{lllrr}
\toprule
  &    &     & Control  & Treatment \\
\midrule
 & Age & Mean & \num{34.81} & \num{34.97}\\
 & Wage & Mean & \num{1503.01} & \num{1769.79}\\
   \midrule
Race & Black & Percent & \num{6.66} & \num{7.57}\\
 & Other & Percent & \num{4.57} & \num{5.08}\\
 & Pardo & Percent & \num{47.74} & \num{56.19}\\
 & White & Percent & \num{41.03} & \num{31.16}\\
   \midrule
Gender & Female & Percent & \num{43.35} & \num{33.99}\\
 & Male & Percent & \num{56.65} & \num{66.01}\\
  \midrule
Education & College Education & Percent & \num{7.34} & \num{6.86}\\
 & High School Education & Percent & \num{59.38} & \num{54.62}\\
 & No High School Diploma & Percent & \num{33.28} & \num{38.52}\\
  \midrule
Industry & Construction & Percent & \num{8.98} & \num{12.30}\\
 & Financial Services & Percent & \num{0.88} & \num{0.77}\\
 & Mining & Percent & \num{0.76} & \num{0.00}\\
 & Public Administration & Percent & \num{0.46} & \num{0.43}\\
 & Rural Activities & Percent & \num{1.49} & \num{6.31}\\
 & Tourism & Percent & \num{7.48} & \num{4.64}\\
   \midrule
 & All & N & \num{169768} & \num{329146}\\
\bottomrule
\end{tabular}
\end{adjustbox}
\end{table}

\end{document}